# The effects of non-tariff measures on agri-food trade: a review and meta-analysis of empirical evidence

*Fabio Gaetano Santeramo*[*], *Emilia Lamonaca*

*University of Foggia (Italy)*

**Abstract**

*The increasing policy interests and the vivid academic debate on non-tariff measures (NTMs) has stimulated a growing literature on how NTMs affect agri-food trade. The empirical literature provides contrasting and heterogeneous evidence, with some studies supporting the 'standards as catalysts' view, and others favouring the 'standards as barriers' explanation. To the extent that NTMs can influence trade, understanding the prevailing effect, and the motivations behind one effect or the other, is a pressing issue.*

*We review a large body of empirical evidence on the effect of NTMs on agri-food trade and conduct a meta-analysis to disentangle potential determinants of heterogeneity in estimates. Our findings show the role played by the publication process and by study-specific assumptions. Some characteristics of the studies are correlated with positive significant estimates, others covary with negative significant estimates. Overall, we found that the effects of NTMs vary across types of NTMs, proxy for NTMs, and levels of details of studies. Not negligible is the influence of methodological issues and publication process.*

[*] Corresponding author: Fabio G. Santeramo, IATRC Member. Email: fabio.santeramo@unifg.it.

# The effects of non-tariff measures on agri-food trade: a review and meta-analysis of empirical evidence

## 1. Introduction

Since the negotiations of the World Trade Organisation (WTO), which have substantially reduced tariffs and fostered global trade, the level of non-tariff measures (NTMs) has remained high and, indeed, has increased over time (Harvey, 1994). The NTMs are policy measures, alternative to tariffs, capable of modifying trade flows (Arita et al., 2017). The growing use of NTMs has led to a less transparent trade policy environment (Fernandes et al., 2017), which calls for a deeper understanding on if and how NTMs influence trade.

The importance of understanding the trade effects of NTMs in agri-food sector is attested by the large and increasing number of papers hosted in top journals (e.g. Hooker and Caswell, 1999; Wilson and Otsuki, 2004; Dawson 2006; Sheldon, 2006; Olper and Raimondi, 2008; Cioffi et al., 2011; Beckman and Arita, 2016). In particular, the number of studies on trade effects of NTMs has significantly increased after 2000 (14 studies available up to 2000, while 140 studies published up to 2017), an upward trend that is parallel to the upward trend of the number of implemented NTMs (from 1.09 million of NTMs notified in 2000, to 4.35 million in 2017). However, the empirical evidence on the issue are heterogeneous and controversial: the magnitude and the direction (positive *versus* negative) of trade effects of NTMs, estimated in literature, are not clear cut. Some issues deserve attention and, in particular, we pose the following questions: what is the prevailing effect of NTMs on global agri-food trade? Do NTMs enhance or impede trade? Which factors influence the heterogeneity in estimates?

These issues have been partly investigated by Cipollina and Salvatici (2008), Li and Beghin (2012), Beghin et al. (2015), and Salvatici et al. (2017). These reviews examine specific categories of NTMs or particular geographic areas: Cipollina and Salvatici (2008) pay attention to trade policies implemented at the border; Salvatici et al. (2017) deepen on the trade measures of the European Union. In addition, the existing reviews are qualitative analyses, exception made for Li and Beghin (2012) who propose a meta-analysis to explain the causes of variation in estimated effects of technical measures on trade of agri-food and manufacturing industries. By analysing a set of 27 papers that are theoretically based on gravity model, Li and Beghin (2012) show that some determinants (e.g. specific agri-food sectors, exclusion of multilateral resistance terms) are associated with trade-impeding effects of technical measures. Some issues, however, are still underinvestigated. For instance, Li and Beghin (2012) do not deepen on the effects of the review process (a major driver in meta-analyses) and on the influence of types and proxy of NTMs, as well as on the aggregation level at which the study is conducted[1].

We review empirical evidence on the effects of NTMs on global trade of agri-food products. We conduct a meta-analysis to conclude on potential determinants of heterogeneity in estimates. In order to complement previous studies, we analyse a larger, and more recent set of empirical researches on the trade effects of NTMs, by following methodological arguments of meta-analysis (Stanley and Jarrell, 1989; Stanley, 2005; Stanley et al., 2008; Stanley and Doucouliagos, 2012; Doucouliagos and Stanley, 2013; Stanley et al., 2013). We explain how magnitude, direction, statistical significance, and accuracy of estimates depends on types of NTMs, proxies used for NTMs, level of detail at which the study is conducted, methodological issues, and publication process. Our analysis complements the existing debate on how, and in which direction, NTMs tend to influence trade.

## 2. The trade effect of non-tariff measures

### *2.1 A theoretical perspective*

Tariffs are protectionist by definition: they undermine the social welfare by crowding out trade (Swinnen, 2016). Non-tariff instruments may be protectionist or competitive for trade: they imply welfare redistributions by addressing market imperfections such as asymmetric information and externalities (Xiong and Beghin, 2014). From a social perspective, while the optimal level of tariffs is zero, determining the optimal level of non-tariff instruments is challenging (Swinnen and Vandemoortele, 2011; Swinnen, 2017) due to the complex relationship linking trade and social effects of non-tariff instruments (Sheldon 2012).

The understanding of non-tariff instruments has changed overtime: as the term 'non-tariff barriers' (NTBs), which emphasises their protectionist scopes (e.g. quotas, export restraints), has been replaced by 'non-tariff measures' (NTMs), in order to emphasise their potential role of hampering or facilitating trade (Grant and Arita, 2017).

According to the definition proposed by UNCTAD (2012), NTMs are policy measures, other than ordinary customs tariffs, that may have economic effects on international trade of goods, changing traded quantities and/or prices. NTMs may also have a corrective role, by reducing asymmetric information (Technical Barriers to Trade, TBTs), mitigating risks in consumption, improving the sustainability of eco-systems (Sanitary and Phytosanitary Standards, SPSs), and influencing the competition and the decision to import or export (non-technical NTMs).

[1] Appendix A.1 provides a detailed comparison with Li and Beghin (2012).

In a small open economy the policymaker sets NTMs on a product category, produced in domestic market and imported from country's trading partners, in order to maximise the domestic welfare: in domestic market, the optimal level of NTMs depends on the trade-off between the marginal utility gain for consumers and the marginal cost for producers. The effects on domestic welfare are influenced by trade strategies of trading partners. Exception made for the case in which the effects on domestic production exactly offset the effects on domestic consumption (Swinnen, 2016), NTMs are capable of influencing trade.

From consumers' perspective, NTMs are socially desirable and provide higher social well-being: by reducing asymmetric information and/or externalities, NTMs enhance consumers' trust, reduce transaction costs and increase consumers' demand (Xiong and Beghin, 2014). The growing demand and the higher costs of implementing NTMs increase the equilibrium price and, as a consequence, the consumption expenditures. The net effect of NTMs on consumers' surplus depends on the magnitude of (positive) utility gain compared to the size of (negative) effect on consumption expenditures: the higher the consumers' utility, the higher the willingness to pay a higher price for products under regulation (Crivelli and Gröschl, 2016; Swinnen, 2016).

From producers' perspective, NTMs imply higher costs of compliance, both fixed costs (e.g. upgrade of practice codes and facilities, acquisition of certificates, conformity in marketing requirements) and variable costs (e.g. prolonged delivery time due to inspection and testing procedures at custom points, rejection of certain shipments, denial of entry of certain shipments) (Xiong and Beghin, 2014; Crivelli and Gröschl, 2016), determining a reduction in profits and supply. The reduced supply increases the equilibrium price and producers' revenue. The net effect on producers' profits depends on the magnitude of (positive) gain in revenue, compared to the size of (negative) implementation costs: the lower the implementation costs, the higher the gain in revenue for products under regulation (Swinnen, 2016).

For exporters, a NTM implemented in the destination country implies higher costs of compliance and a higher import price. If the difference between import price pre- and post-NTM is greater (smaller) than the difference between domestic price pre- and post-NTM, domestic producers face smaller (greater) implementation costs and obtain greater (lower) profits than foreign producers. The NTM acts as barrier (catalyst) for trade if it reduces (increases) domestic imports (Swinnen, 2017).

### *2.2 An empirical perspective*

The empirical literature on non-tariff measures (NTMs) and trade provides mixed evidence. Several studies suggest that NTMs hamper trade (e.g. Peterson et al., 2013; Dal Bianco et al., 2016), others conclude that they foster trade (e.g. Cardamone, 2011), and numerous studies show mixed effects of NTMs on trade (e.g. Xiong and Beghin, 2011; Beckman and Arita, 2016). Such heterogeneity may be explained by the characteristics of empirical studies, which differ in terms of design and methods.

Only few studies provide a general assessment of the trade effects of NTMs: a remarkable case is Hoeckman and Nicita (2011) who suggest that NTMs are major frictions to trade of agri-food products. Differently, most of empirical studies are partial and focused on NTM-, product-, or country-specific case studies.

Different types of NTMs may lead to different empirical results. Different interventions may have different effects on trade (Schlueter et al., 2009). In addition, the lower the aggregation of NTMs under investigation, the crisper the policy implication for addressed issues (Li and Beghin, 2012). Accordingly, in literature we find that Technical Barriers to Trade (TBTs) tend to be catalysts for trade (e.g. de Frahan and Vancauteren, 2006), whereas Sanitary and Phytosanitary Standards (SPSs) show mixed evidence (e.g. Schlueter et al., 2009; Jayasinghe et al., 2010; Crivelli and Gröschl, 2016). Divergences may be due to the peculiarity of the SPSs, which may have either: "*a substantial positive impact [… or] a significant negative impact.*" (Schlueter et al., 2009, p. 1489). The Maximum Residue Levels (MRLs) tend to act as barrier to trade (e.g. Otsuki et al., 2001 a, b; Chen et al., 2008; Ferro et al., 2015).

The effects of NTMs may be also sector- and/or product- specific. For instance, NTMs are likely to be trade-impeding for seafood products (e.g., Anders and Caswell 2009), meat (Wilson et al., 2003), fruits and vegetables, cereals and oil seeds (e.g., Otsuki et al., 2001a, b; Scheepers et al., 2007; Drogué and DeMaria 2012). Vice-versa, trade of fats and oils seems not affected by beyond-the-border policies (e.g. Xiong and Beghin 2011).

In addition, countries involved in empirical analyses may determine specific geo-economic patterns of NTMs . In this regard, studies that investigate the impacts of NTMs implemented by developed countries against developing countries are frequent in literature, and tend to show negative effects on trade performances of developing countries (e.g. Anders and Caswell, 2009; Disdier and Marette, 2010). Vice-versa, NTMs may have either negative (e.g. Yue and Beghin, 2009) or positive effects (de Frahan and Vancauteren, 2006) on trade among developed countries. Differently, NTMs tend to limit trade among developing countries (Melo et al., 2014).

Other sources of heterogeneity  may be related to the variety of methodological and empirical approaches we find in literature. An example are the different proxies used to measure NTMs: some methodologies include inventory measures (e.g. dummy or count variables, frequency index, coverage ratio, prevalence score), computation of price gaps, and the estimation of *ad valorem* *equivalents* (AVEs) (Gourdon, 2014). In literature, the effects on trade tend to be negative if NTMs are proxied by AVE (e.g. Olper and Raimondi, 2008; Arita et al., 2017), or by frequency index and/or coverage ratio (e.g. Jongwanich, 2009; Fernandes et al., 2017). Differently, if NTMs are proxied by dummy or count variables, the results may be either positive (e.g. Cardamone, 2011; Shepherd and Wilson, 2013) or negative (e.g. Peterson et al., 2013; Dal Bianco et al., 2016).

Lastly, different types of data do matter. For instance, both Schlueter et al. (2009) and Beckman and Arita (2016) estimate the effect of SPSs on trade of meat between developed countries: the formers use data aggregated at HS-4 digit and estimate a positive effect on trade; the latters use data aggregated at HS-6 digit and highlight a negative effect on trade.

Due to the large heterogeneity in trade effects of NTMs, a systematic assessment of potential determinants of these effects is worth.

## 3. The meta-analytical approach

The heterogeneity characterises all economic researches (Havránek, 2010): the empirical literature on trade effects of non-tariff measures (NTMs) is not exempt. Since the pioneering work of Stanley and Jarrell (1989), economic meta-analyses (MAs) aim at assigning a pattern to such heterogeneity.

The MA is an econometric approach that allows to combine and summarise evidence from different but comparable empirical studies, and to explain the wide variation of results across studies (Stanley et al., 2008). This analytic technique allows to test competing theories and to synthesize empirical estimates (Stanley and Doucouliagos, 2012). Doucouliagos and Stanley (2013) suggest that theoretical competition may shape the distribution of reported empirical findings and worsen publication selectivity. The publication selection is a main concern of MA: it may bias estimates and create heterogeneity across studies, undermining the validity of inferences and policy implications (Stanley et al., 2008; Doucouliagos and Stanley, 2013).) Publication selection biases may concern the propensity of academics towards a particular direction of results (i.e. negative or positive estimates) (type I bias), or may occur if statistically significant results are treated more favourably, thus are more likely to be published (type II bias) (Stanley, 2005; Stanley and Doucouliagos, 2014).

The MA is becoming more and more popular in economics: for instance, it has been applied to the price elasticity of demand (Böcker and Finger, 2017), the calorie-income elasticity (Santeramo and Shabnam, 2015), and to food safety (Xavier et al., 2014). It has been also used to investigate international trade: Rose and Stanley (2005) and Havránek (2010) analyse the effect of currency unions on trade; Disdier and Head (2008) examine potential causes of variation in distance effect on bilateral trade; Cipollina and Salvatici (2010) investigate the impact of preferential trade agreements on intra-bloc trade; Li and Beghin (2012) explain variations in estimated trade effects of technical barriers to trade. We focus on trade effects of NTMs in the agri-food sector.

### *3.1 Literature searching criteria and selection process*

During the last twenty-five years the trend of (theoretical, 32%, and empirical, 68%) papers on non-tariff measures (NTMs) has been exponential (figure 1).

Figure 1. Trends of studies on trade effects of NTMs and notified NTMs in the agri-food sector.

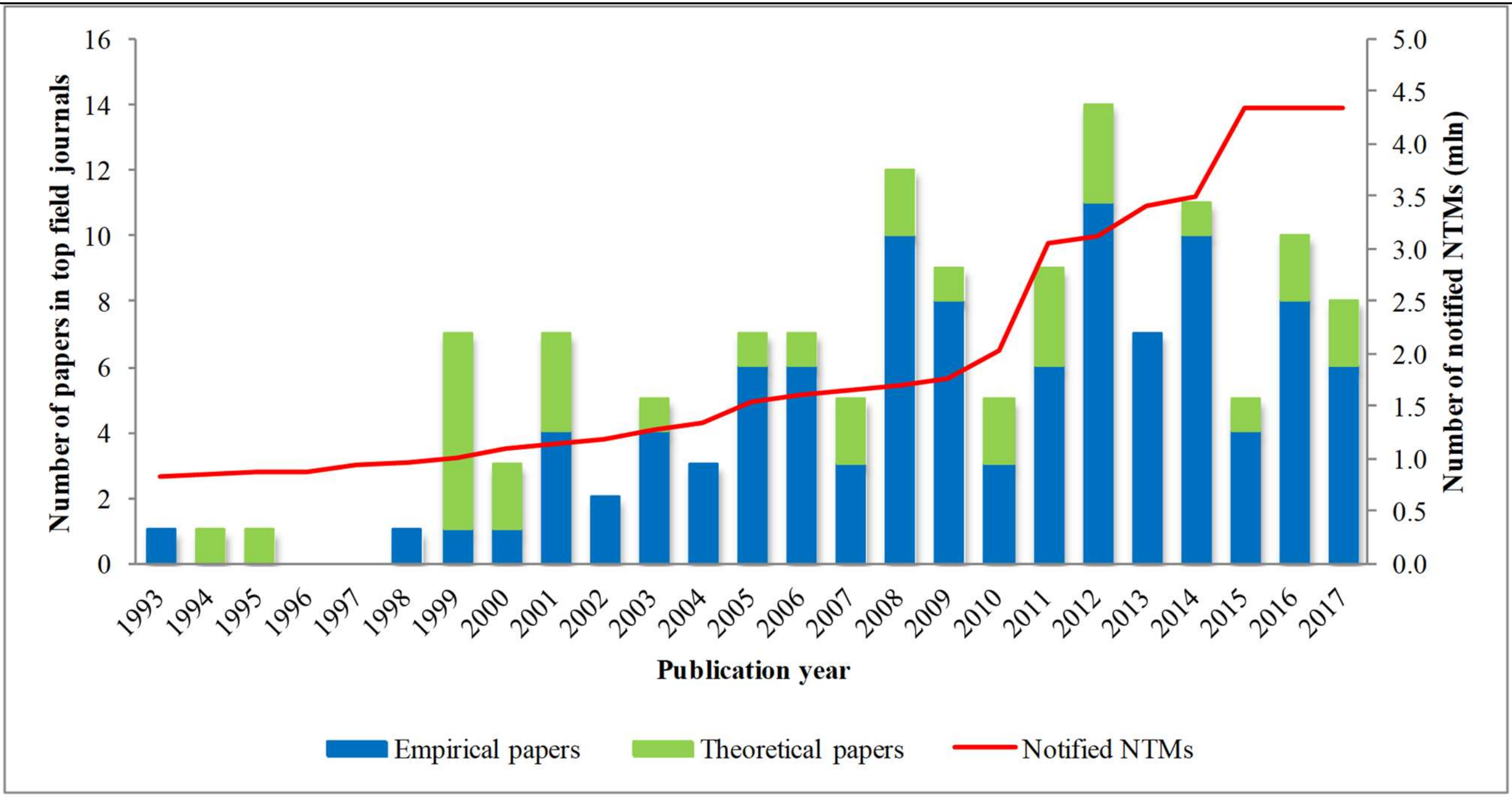

Source: elaboration on UNCTAD (2017), TRAINS NTMs: The Global Database on Non-Tariff Measures.

Note: We refer to top field journals (listed in Appendix A.1).

We systematically reviewed the literature following the guidelines provided in Stanley et al. (2013). We searched studies in bibliographic databases: Scopus, Web of Science, and JSTOR provided access to multidisciplinary information from prestigious, high impact research journals; RePEc, IATRC, AgEcon Search, and Google Scholar allowed us to cover grey literature[2] (i.e. working papers and conference proceedings); repositories of specific peer-reviewed journals[3] and papers

[2] In line with other analyses on trade issues (e.g. Disdier and Head, 2008; Cipollina and Salvatici, 2010; Li and Beghin, 2012), we include working papers and conference proceedings to identify publication selection. The journal prestige may be a source of publication bias (Santeramo and Shabnam, 2015). In order to avoid double counting, we include working papers and conference proceedings that do not correspond to revised versions published in peer-reviewed journals.

[3] We consider the following journals: European Review of Agricultural Economics, American Journal of Agricultural Economics, Journal of Agricultural Economics, Agricultural Economics, Australian Journal of Agricultural and Resource Economics, Canadian Journal of Agricultural Economics, Applied Economics Policy Perspective, World Bank Economic Review, World Development,

cross-references traced back further works. The search was carried out in August 2017 and was limited to researches published in a period ranging from 1990 to 2017.

We used keywords such as ‘trade’ and ‘agri-food trade’, combined with other terms: ‘non-tariff measure/non-tariff barrier’, ‘technical barrier to trade’, ‘sanitary and phytosanitary standard’, ‘maximum residue level’, ‘specific trade concern’. We identified 155 studies. Subsequently, each paper has been reviewed in depth, so to limit the analysis to papers that assess the trade effects of NTMs: we excluded theoretical studies, and papers that do not provide comparable empirical results[4]. The final sample includes 62 papers (47 published in peer-reviewed journals, and 15 from grey literature), 1,362 observations (point estimates of trade effects of measures, ETEMs) and 1,213 estimated t-statistics[5] (table 1)[6].

---

Agribusiness, Journal of Development Economics, Journal of Development Studies, China Agricultural Economic Review, German Journal of Agricultural Economics.

[4] The appendix A.2 provides a list of excluded studies.

[5] We have 149 missing values for t-statistics due to the lack, in some papers, of standards errors and t-values.

[6] Appendix A.3 provides descriptive statistics for each papers included in the sample.

Table 1. Papers included in the empirical analysis.

| Authors | Publication year | Publication outlet[a] | Type of NTM[b] | Main effect of NTMs on trade |
|---|---|---|---|---|
| Anders and Caswell | 2009 | AJAE | Other | Negative |
| Arita, Beckman, and Mitchell | 2017 | FP | TBT; SPS | Negative |
| Babool and Reed | 2007 | EAAE CP | MRL | Negative |
| Beckman and Arita | 2016 | AJAE | SPS; Other | Mixed effects |
| Cardamone | 2011 | ERAE | Other | Positive |
| Chen, Yang, and Findlay | 2008 | RWE | MRL | Negative |
| Chevassus-Lozza, Latouche, Majkovic, and Unguru | 2008 | FP | TBT; SPS; MRL; Other | Mixed effects |
| Crivelli and Gröschl | 2016 | WE | SPS | Mixed effects |
| Dal Bianco, Boatto, Caracciolo, and Santeramo | 2016 | ERAE | TBT; SPS | Negative |
| de Frahan and Vancauteren | 2006 | ERAE | TBT | Positive |
| Disdier and Fontagné | 2008 | EAAE CP | Other | Negative |
| Disdier and Marette | 2010 | AJAE | MRL | Negative |
| Disdier, Fekadu, Murillo, and Wong | 2008a | ICTSD WP | TBT; SPS | Negative |
| Disdier, Fontagné, and Mimouni | 2008b | AJAE | TBT; SPS | Negative |
| Drogué and DeMaria | 2012 | FP | MRL; Other | None |
| Essaji | 2008 | JIE | NTM | Negative |
| Fernandes, Ferro, and Wilson | 2017 | WBER | MRL | Negative |
| Ferro, Otsuki, and Wilson | 2015 | FP | MRL | Negative |
| Ferro, Wilson, and Otsuki | 2013 | World Bank WP | MRL | Negative |
| Fontagné, Mayer, and Zignago | 2005 | CJE | NTM | Negative |
| Gebrehiwet, Ngqangweni, and Kirsten | 2007 | Agrekon | MRL | Negative |
| Harrigan | 1993 | JIE | NTM | Negative |
| Hoekman and Nicita | 2011 | WD | NTM | Negative |
| Jayasinghe, Beghin, and Moschini | 2010 | AJAE | SPS | Negative |
| Jongwanich | 2009 | FP | SPS | Negative |
| Kareem | 2014a | CP | SPS | Mixed effects |
| Kareem | 2014b | EUI RSCAS WP | SPS | Mixed effects |

Table 1. (Continued).

| Authors | Publication year | Publication outlet[a] | Type of NTM[b] | Main effect of NTMs on trade |
|---|---|---|---|---|
| Kareem | 2014c | WP | SPS | Mixed effects |
| Kareem | 2016a | ITJ | SPS | Mixed effects |
| Kareem | 2016b | JAD | SPS | Mixed effects |
| Kareem | 2016c | JCM | Other | Mixed effects |
| Kareem, Brümmer, and Martinez-Zarzoso | 2015 | Global Food WP | MRL; Other | Negative |
| Melo, Engler, Nahuehual, Cofre, and Barrena | 2014 | WD | TBT; SPS; MRL; Others | Mixed effects |
| Moenius | 2004 | WP | Other | Mixed effects |
| Moenius | 2006 | IATRC CP | Other | Mixed effects |
| Munasib and Roy | 2013 | IAAE CP | MRL | Mixed effects |
| Nardella and Boccaletti | 2003 | AAEA CP | NTM; TBT; SPS | Negative |
| Nardella and Boccaletti | 2004 | AAEA CP | SPS | Mixed effects |
| Nardella and Boccaletti | 2005 | AAEA CP | TBT | Positive |
| Nguyen and Wilson | 2009 | SAEA CP | SPS | Negative |
| Olper and Raimondi | 2008 | JAE | NTM; Other | Negative |
| Otsuki, Wilson, and Sewadeh | 2001a | ERAE | MRL | Negative |
| Otsuki, Wilson, and Sewadeh | 2001b | FP | MRL | Negative |
| Péridy | 2012 | ERI | NTM | Negative |
| Peterson, Grant, Roberts, and Karov | 2013 | AJAE | SPS | Negative |
| Saitone | 2012 | Agribusiness | Other | None |
| Scheepers, Jooste, and Alemu | 2007 | Agrekon | MRL | Negative |
| Schlueter, Wieck, and Heckelei | 2009 | AJAE | SPS | Mixed effects |
| Schuster and Maertens | 2013 | FP | Other | Negative |
| Shepherd and Wilson | 2013 | FP | Other | Mixed effects |
| Shepotylo | 2016 | MP | TBT; SPS; STC | None |
| Sun, Huang, and Yang | 2014 | CAER | MRL | None |
| Tran, Nguyen, and Wilson | 2014 | Agribusiness | MRL | Mixed effects |
| Vollrath, Gehlhar, and Hallahan | 2009 | JAE | Other | Mixed effects |

Table 1. (Continued).

| Authors | Publication year | Publication outlet[a] | Type of NTM[b] | Main effect of NTMs on trade |
|---|---|---|---|---|
| Wei, Huang, and Yang | 2012 | CER | MRL | Negative |
| Wilson and Otsuki | 2003 | JIE | MRL | Negative |
| Wilson and Otsuki | 2004 | FP | MRL; Other | Mixed effects |
| Wilson, Otsuki, and Majumdar | 2003 | JITED | MRL | Negative |
| Winchester et al. | 2012 | WE | TBT; SPS; MRL | Negative |
| Xiong and Beghin | 2011 | ERAE | MRL | Mixed effects |
| Xiong and Beghin | 2014 | EI | MRL | Mixed effects |
| Yue and Beghin | 2009 | AJAE | TBT; SPS | Negative |

[a]Acronyms are as follows: American Journal of Agricultural Economics (AJAE), Food Policy (FP), European Association of Agricultural Economists (EAAE), Conference Proceeding (CP), European Review of Agricultural Economics (ERAE), Review of World Economics (RWE), World Economics (WE), International Centre for Trade and Sustainable Development (ICTSD), Journal of International Economics (JIE), World Bank Economic Review (WBER), Working Paper (WP), Canadian Journal of Economics (CJE), World Development (WD), European University Institute Robert Schuman Centre for Advanced Studies (EUI RSCAS), International Trade Journal (ITJ), Journal of African Development (JAD), Journal of Commodity Markets (JCM), International Agricultural Trade Research Consortium (IATRC), International Association of Agricultural Economists (IAAE), American Agricultural Economics Association (AAEA), Southern Agricultural Economics Association (SAEA), Journal of Agricultural Economics (JAE), Economic Research International (ERI), Marine Policy (MP), China Agricultural Economic Review (CAER), China Economic Review (CER), Journal of International Trade & Economic Development (JITED), Economic Inquiry (EI).

[b]Acronyms are as follows: NTMs stands for ‘Non-Tariff Measures’ (as general category), TBT stands for ‘Technical Barrier to Trade’, SPS stands for ‘Sanitary and Phytosanitary Standard’, MRL stands for ‘Maximum Residue Level’, STC stands for ‘Specific Trade Concern’. ‘Other’ includes measures not involved in previous categories, such as quality and quantity restrictions, Hazard Analysis and Critical Control Points (HACCP), private standards, voluntary standards.

*3.2 Empirical model*

The heterogeneity in the estimated trade effects of measures (ETEMs) is likely to depend on publication selection and characteristics of empirical studies. The t-statistics of ETEMs[7] ($\hat{t}$) are regressed on the precision of the estimates (i.e. the inverse of the estimated standard error, $\frac{1}{\hat{\sigma}}$), on *J* regressors related to the characteristics of the study ($\chi_j$), and on *K* regressors related to potential publication selection ($Z_k$):

$$\hat{t} = \alpha_0 + \alpha_1 \frac{1}{\hat{\sigma}} + \sum_{j=1}^{J} \beta_j \frac{\chi_j}{\hat{\sigma}} + \sum_{k=1}^{K} \gamma_k Z_k + \varepsilon \qquad (1)$$

In line with Stanley et al. (2008), we include both variables ($\chi_j$) that are likely to influence the estimates, but that are uncorrelated with the likelihood of acceptance, as well as variables ($Z_k$) that may influence the likelihood of acceptance for publication, but should not be informative on the estimates. The constant term ($\alpha_0$) collects potential information on the publication selection that are not directly included in the model (Stanley and Jarrell, 1989). The error term ($\varepsilon_i$) is assumed to be independently and identically distributed (i.i.d.).

We estimate model in equation (1) through a robust regression technique to mitigate potential problems related to outliers and influential data points (Belsley et al., 1980). Influential data points are likely to exist in our sample because we use multiple estimates from the same study (that are likely to be correlated). Coefficients of the robust regression allow us to infer on the magnitude of ETEMs.

In order to determine which drivers may explain the direction (positive or negative) of statistically significant ETEMs, we use a Multinomial Logit (MNL) model: the dependent variable is categorical ($Y_{MNL}$)[8] and it allows us to classify the ETEMs as negative (t-statistic lower than −1.96), not significant (t-statistic between −1.96 and 1.96), or positive (t-statistic higher than 1.96):

$$Y_{MNL} = \begin{cases} -1 \; if \; \widehat{t} \leq -1.96 \\ 0 \; if \; -1.96 < \widehat{t} < 1.96 \\ 1 \; if \; \widehat{t} \geq 1.96 \end{cases} \qquad (2)$$

By substituting the equation (2) in (1), we derive a system of two equations:

---

[7] We use estimated t-statistics instead of ETEMs to avoid problems of heteroschedasticity (Stanley, 2001).

[8] Differently from Li and Beghin (2012), we classify negative significant ETEMs as '-1' (instead of '1'), not significant ETEMs as '0' (instead of '2'), and positive significant ETEMs as '1' (instead of '3').

$$\begin{cases} \ln\left(\frac{Pr(Y_{MNL}=-1)}{Pr(Y_{MNL}=0)}\right) = \alpha_0 + \alpha_1 \frac{1}{\hat{\sigma}} + \sum_{j=1}^{J} \beta_j \frac{\chi_j}{\hat{\sigma}} + \sum_{k=1}^{K} \gamma_k Z_k + \varepsilon \\ \ln\left(\frac{Pr(Y_{MNL}=1)}{Pr(Y_{MNL}=0)}\right) = \alpha_0 + \alpha_1 \frac{1}{\hat{\sigma}} + \sum_{j=1}^{J} \beta_j \frac{\chi_j}{\hat{\sigma}} + \sum_{k=1}^{K} \gamma_k Z_k + \varepsilon \end{cases} \quad (3)$$

where $\ln\left(\frac{Pr(Y_{MNL}=-1)}{Pr(Y_{MNL}=0)}\right)$ and $\ln\left(\frac{Pr(Y_{MNL}=1)}{Pr(Y_{MNL}=0)}\right)$ are the logarithms of the probability of having, respectively, negative and significant (rather than not significant) or positive and significant (rather than not significant) ETEMs.

A Probit model is adopted to explain the drivers for statistical significance, and a Tobit model is used to deepen on the magnitude of the estimated t-statistics (accuracy of ETEMs). The dependent variable of the Probit model is a dummy variable equal to 1 if the *i*-th ETEMs is statistically significant (t-statistics lower than −1.96, or higher than 1.96), and 0 otherwise:

$$Y_{Probit} = \begin{cases} 1 \; if \; \hat{t_\iota} \leq -1.96 \; or \; \hat{t_\iota} \geq 1.96 \\ 0 \text{ otherwise} \end{cases} \quad (4)$$

The dependent variable of the Tobit model is a continuous variable equal to the t-statistics of ETEMs ($\hat{t}$), if it is larger than the threshold value (1.96 in absolute value), and 0 otherwise:

$$Y_{Tobit(-1.96)} = \begin{cases} \hat{t_\iota} \; if \; \hat{t_\iota} \leq -1.96 \\ 0 \; if \; \hat{t_\iota} > -1.96 \end{cases} \quad \text{and} \quad Y_{Tobit(1.96)} = \begin{cases} \hat{t_\iota} \; if \; \hat{t_\iota} \geq 1.96 \\ 0 \; if \; \hat{t_\iota} < 1.96 \end{cases} \quad (5)$$

In the Tobit model, both right- and left-censored, positive coefficients imply greater t-statistics and, thus, less accurate ETEMs; vice-versa for negative coefficients.

In line with Li and Beghin (2012), we use a robust estimator of the clustered error structure to estimate MNL, Probit, and Tobit models. We assume independence among clusters (i.e. papers), and dependence among observations within each cluster (i.e. ETEMs of the same paper).

### *3.3 Description of covariates*

Our model includes covariates related to the characteristics of the study ($\chi_j$) and to the publication selection ($Z_k$), to explain heterogeneity in estimated effects of non-tariff measures (NTMs).The set of covariates related to the characteristics of the study allows us to control for types of NTMs, proxies for NTMs, and the level of detail of the study. In particular, specific dummy variables account for types of measure (i.e. Technical Barrier to Trade, TBT, Sanitary and Phytosanitary Standard, SPS, Maximum Residue Level, MRL[9]).

Further dummies are used to proxy the intensive and the extensive margins of the NTM: *ad valorem equivalent*, or AVE, proxies the intensive margins by capturing the degree of protectionism of the NTMs (i.e. how much NTMs affect trade); dummy variables and indices proxy the extensive margins (e.g. existence or not of NTMs).

We also use dummy variables to identify the level of details of the study: a dummy controls for the geo-economic affinity of countries that implement NTMs (i.e. reporters) and of countries affected by NTMs (i.e. partners): we classify reporters and partners into Northern (Developed Economies) and Southern (Developing Economies and Economies in transition) countries, according to the classification of the United Nations (2017). Other dummies control for the level of product aggregation (according to the Harmonised System[10]), and for the specific product category under investigation.

We control for potential publication selection: some covariates are related to methodological issues, others to the publication process. In particular, we control for the adoption of fixed effects to account for multilateral trade resistance terms in gravity models, and for the treatment of zero trade flows. The zero trade flows problem is a common issue in studies based on the gravity equation; our sample includes several papers (87%) based on the gravity model.

As for the publication process, we account for the prestige of the publication outlet, and for grey literature with specific dummies: one dummy controls for papers published in Q1 journals (according to the rank provided by Scimago Journal & Country Rank at the date of publication for the subject area 'Economics and Econometrics') and one dummy accounts for working papers. Furthermore, we use a dummy variable to control for the presence of more than one article published by the same author. Table 2 lists the covariates.

[9] Countries frequently fix MRLs, as an alternative to SPSs, in order to ensure safe imports. The requirements on MRLs are not set in the WTO consultations, but they may be assimilated to the SPS A200 that sets the tolerance limits for residues and imposes a restricted use of certain substances in food and feed (UNCTAD, 2012). Due to these considerations, we distinguish SPSs and MRLs in separate categories.

[10] Commonly known as Harmonised System (HS), the Harmonised Commodity Description and Coding System is the internationally standardised system of names and codes used to classify traded products. We consider four level of aggregation: HS-2 digit, which corresponds to a Chapter (e.g. 09 - Coffee, Tea, Mate and Spices), HS-4 digit, which corresponds to a Heading (e.g. 0901 - Coffee, whether or not roasted or decaffeinated; Coffee husks and skins; Coffee substitutes containing coffee), HS-6 digit, which corresponds to Sub Heading (e.g. 090121 - Coffee, roasted, not decaffeinated), and HS-8 digit which corresponds to Subheading determining duty (e.g. 09012100 - no distinction with respect to 090121).

Table 2. Description of covariates and basic statistics.

| Covariates | Description | Type of variable | Detail | Mean |
|---|---|---|---|---|
| t-statistic | t-statistics related to ETEMs, estimated in literature | Continuous | | -0.10 |
| Standard error | Standard error related to ETEMs, estimated in literature | Continuous | | 0.91 |
| TBT | Standard is a Technical Barrier to Trade | Dummy | 1 if TBT (0 otherwise) | 0.15 |
| SPS | Standard is a Sanitary and Phytosanitary Standard | Dummy | 1 if SPS (0 otherwise) | 0.27 |
| MRL | Standard is a Maximum Residue Level | Dummy | 1 if MRL (0 otherwise) | 0.18 |
| AVE | *Ad valorem equivalent* (AVE) used to proxy standard | Dummy | 1 if AVE (0 otherwise) | 0.06 |
| Dummy variable | Dummy variable used to proxy standard | Dummy | 1 if dummy variable (0 otherwise) | 0.35 |
| Index | Frequency index or coverage ratio used to proxy standard | Dummy | 1 if index (0 otherwise) | 0.10 |
| N-N | North-North | Dummy | 1 if reporter and partner are developed countries (0 otherwise) | 0.28 |
| N-S | North-South | Dummy | 1 if reporter is a developed country and partner is a developing country (0 otherwise) | 0.32 |
| HS-2 digit | Product aggregated at 2 digits of Harmonised System | Dummy | 1 if product is aggregated at HS-2 digit (0 otherwise) | 0.26 |
| HS-4 digit | Product aggregated at 4 digits of Harmonised System | Dummy | 1 if product is aggregated at HS-4 digit (0 otherwise) | 0.36 |
| Meat | Product category under investigation is meat | Dummy | 1 if product category is meat (0 otherwise) | 0.10 |
| Dairy | Product category under investigation is dairy produce | Dummy | 1 if product category is dairy produce (0 otherwise) | 0.05 |
| Cereal | Product category under investigation is cereals | Dummy | 1 if product category is cereal (0 otherwise) | 0.09 |
| Oilseed | Product category under investigation is oil seeds and oleaginous fruits | Dummy | 1 if product category is oilseeds (0 otherwise) | 0.13 |
| F&O | Product category under investigation is animal or vegetable fats and oils | Dummy | 1 if product category is fats and oils (0 otherwise) | 0.05 |
| Beverage | Product category under investigation is beverage | Dummy | 1 if product category is beverage (0 otherwise) | 0.04 |
| Country-pair f.e. | Country-pair fixed effects | Dummy | 1 if country-pair fixed effects are used (0 otherwise) | 0.07 |
| Time f.e. | Time fixed effects | Dummy | 1 if time fixed effects are used (0 otherwise) | 0.16 |
| Product f.e. | Product/industry/sector fixed effects | Dummy | 1 if product fixed effects are used (0 otherwise) | 0.04 |
| Zero trade | Treatment of zero trade flows | Dummy | 1 if zero trade flows are treated (0 otherwise) | 0.48 |
| Q1 | Peer-reviewed journal in the 1st quartile of Scimago Journal & Country Rank (SJR) at date of publication and subject area 'Economics and Econometrics' | Dummy | 1 if published in Q1 (0 otherwise) | 0.42 |
| WP | Gray literature: working paper | Dummy | 1 if published as WP (0 otherwise) | 0.14 |
| Authors | Paper co-authored by academics experienced on the issue | Dummy | 1 if experienced author (0 otherwise) | 0.61 |

### *3.3.1 Collinearity diagnostics*

Our empirical model involves several dichotomous variables: potential collinearity may arise and confound estimation results. In order to test for collinearity, we use the variance inflation index (VIF) and the condition number: they do not exceed the threshold values 10 and 15, respectively, to avoid problems of multicollinearity (Belsley et al., 1980).

As preliminary analysis, the correlation matrix shows that the correlation between the coefficients for 'Publication bias' and 'SPS', 'N-S' and 'SPS', for 'F&O' and 'Oilseed', and for 'Beverage' and 'Meat' is remarkably high (more than 0.80). Diagnostic outputs suggest possible strong collinearity between the same variables. We dropped the covariates with the relative higher VIF: 'Publication bias', 'Oilseed', 'N-S', and 'Beverage'. Collinearity diagnostics without the problematic covariates show no additional problems.

## 4. Results and discussion

### *4.1 Description of the sample*

Our sample reflects the heterogeneity that is also observed in the literature. Papers supporting the 'standards as barriers' view are the majority (34 papers), while papers that conclude on the 'standards as catalysts' view are in limited number (3 papers). Several papers (21) papers propose mixed evidence.

Several studies contain multiple ETEMs: in order to keep important information, we include all available evidence (Jeppensen et al., 2002), rather than opting for the preferred estimate method, or for a synthesis of the estimates (Card and Krueger, 1995; Stanley, 2001; Rose and Stanley, 2005).

Table 2 describes the estimated trade effects of measures (ETEMs): 56% ETEMs (759 point estimates) are negative; 44% (605 point estimates) are non-negative (positive or zero). A majority of estimates (61%) are statistically significant, with 508 point estimates (37%) being negative and 325 point estimates (24%) being positive. The mean and median values of (total) ETEMs are, respectively, -0.30 and -0.05 (confidence interval ranges from -3.33 to 2.73).

Table 3. Descriptive statistics of the estimated trade effects of measures (ETEMs).

| ETEMs | Min | Max | Median | Mean | Std. Dev. | C.I.[a] | Obs.[b] |
|---|---|---|---|---|---|---|---|
| Total | -38.540 | 54.140 | -0.048 | -0.305 | 3.039 | [-3.334; 2.734] | 100% |
| Non-negative (positive or zero) | 0.000 | 54.140 | 0.280 | 1.031 | 2.966 | [-1.935; 3.997] | 44% |
| Negative | -0.001 | -38.540 | -0.456 | -1.37 | 2.653 | [-4.023; 1.283] | 56% |
| Significant | -38.540 | 18.105 | -0.160 | -0.598 | 3.151 | [-3.749; 2.553] | 61% |
| Significant non-negative (positive or zero) | 0.000 | 18.105 | 0.580 | 1.263 | 2.296 | [-1.033; 3.559] | 24% |
| Significant negative | -38.540 | -0.004 | -0.734 | -1.789 | 3.047 | [-4.836; 1.258] | 37% |
| Not significant | -12.920 | 54.140 | 0.004 | 0.155 | 2.795 | [-2.640; 2.950] | 39% |

Notes: In the sample, only two observations are equal to zero.

[a] Confidence interval (C.I.) ranges between mean minus standard deviation (minimum) and mean plus standard deviation (maximum).

[b] Percentages computed on the total number of observations (1,364).

Figure 2 presents the distribution (boxplots) and the kernel densities of total, positive, and negative ETEMs[11]: half of the statistically significant ETEMs (680 observations out of 1,364) ranges between median values of (statistically significant) negative ($Me_{Neg.}$ = -0.42) and of (statistically significant) positive ($Me_{Pos.}$ = 0.34) observations (figure 2, panel (i)). The distribution of total ETEMs is bimodal with one negative peak ($Mo_{Neg.}$ = -0.21) and one positive peak ($Mo_{Pos.}$ = 0.02). Negative and positive ETEMs are almost equally dispersed (the standard deviations of negative, $\sigma_{Neg.}$ = 2.65, and positive ETEMs, $\sigma_{Pos.}$ = 2.97 are close) (figure 2, panel (ii)).

[11] The distribution and kernel density estimated in figure 2 refer to a subsample which ranges between the $10^{th}$ and the $95^{th}$ percentiles.

Figure 2. Estimated trade effect of measures (ETEMs) arranged by direction.

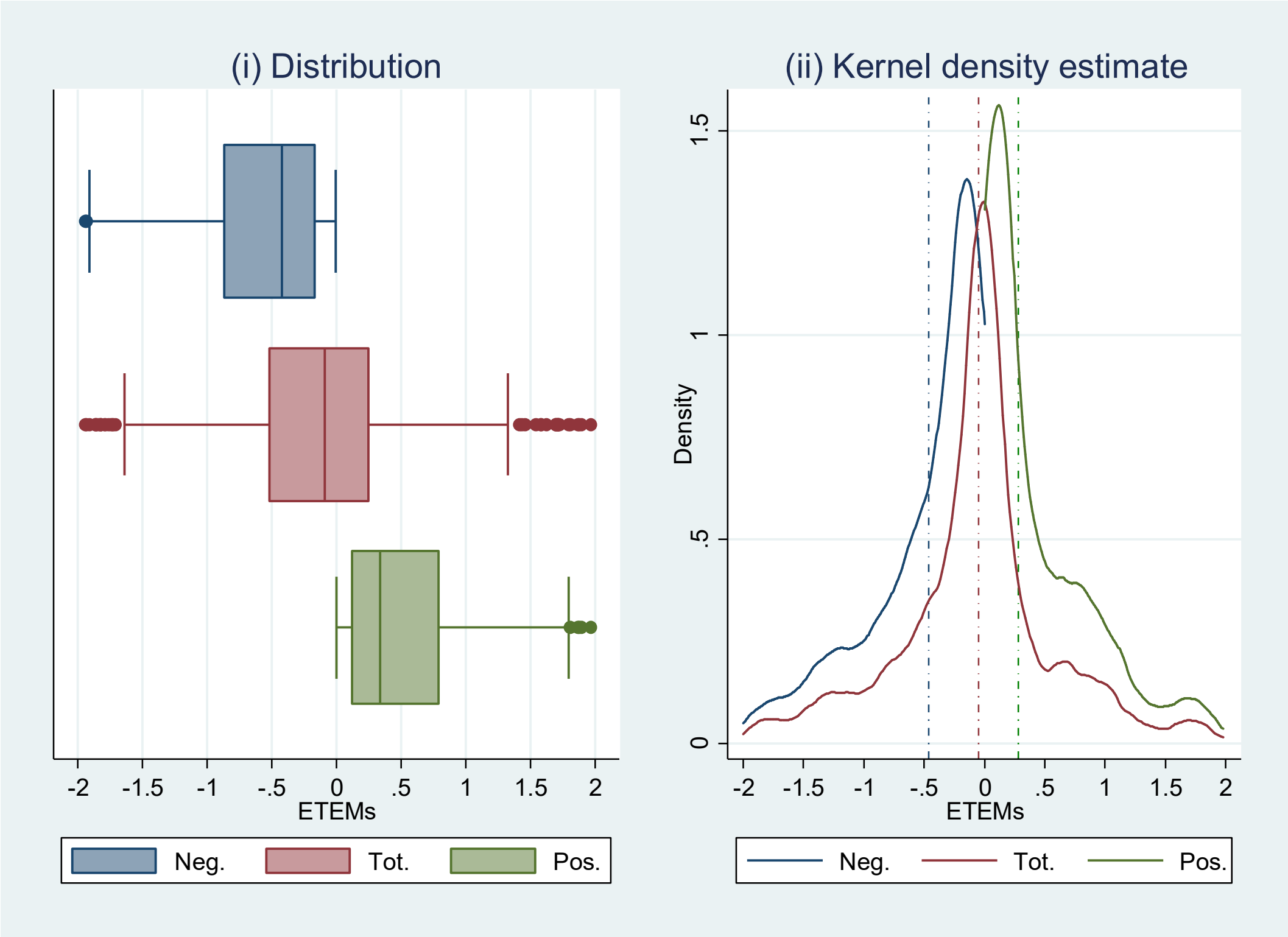

Notes: In panel (i), distributions of ETEMs are on statistically significant observations within the 10th and the 95th percentiles. Horizontal lines within boxes are median values (Me) (i.e. $Me_{Neg.}$ = -0.42, $Me_{Tot.}$ = -0.16, $Me_{Pos.}$ = 0.34). In panel (ii), the estimated densities for ETEMs are computed removing observations which exceed the 10th and the 95th percentiles. Dashed lines are median values (Me) computed on total observations (i.e. $Me_{Tot.}$ = -0.05, $Me_{Pos.}$ = 0.28, $Me_{Neg.}$ = -0.46).

Several variables seem correlated with the magnitude and the direction of NTMs' trade effects (e.g. geo-economic areas, type of NTMs). Figure 3 shows that the ETEMs differ across geo-economic areas. Papers that investigate North-North or North-South trade (40 papers, 822 point estimates) are in larger number with respect to papers that analyse South-North or South-South trade (3 papers, 76 point estimates) (figure 3, panel (i)). The variability of ETEMs across papers that consider North-North ($\sigma_{N\text{-}N}$ = 3.19) or North-South ($\sigma_{N\text{-}S}$ = 4.17) trade is larger compared to that of papers for South-South trade ($\sigma_{S\text{-}S}$ = 0.68). However, the median of ETEMs is lower for North-South ($Me_{N\text{-}S}$ = -0.10) trade than for North-North ($Me_{N\text{-}N}$ = -0.01) trade (figure 3, panel (ii)).

Figure 3. Estimated trade effect of measures (ETEMs) arranged by geo-economic areas.

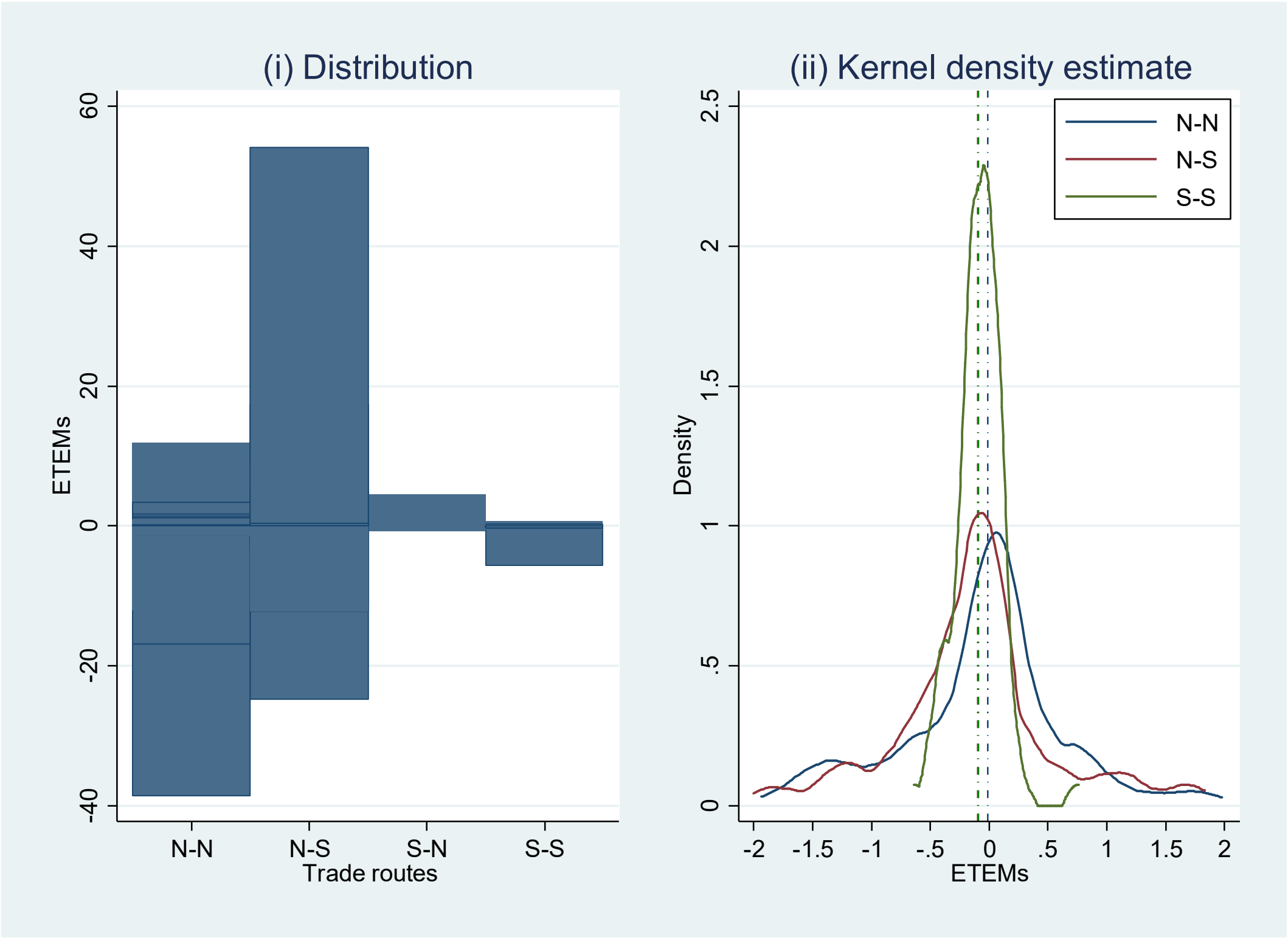

Notes: Geo-economic areas are as follows: N-N stands for 'North-North', N-S stands for 'North-South', S-N stands for 'South-North', S-S stands for 'South-South', where the formers are countries imposing NTMs (reporters) and the latters are countries affected by NTMs (partners). Reporters and partners are classified into North (Developed Economies) and South (Developing Economies and Economies in transition), according to the country classification proposed by the United Nations (2017). In panel (ii), the estimated densities for ETEMs are computed removing observations which exceed the $10^{th}$ and the $95^{th}$ percentiles. Dashed lines are median values (Me) computed on total observations (i.e. $Me_{Tot.}$ = -0.05, $Me_{N\text{-}N}$ = -0.01, $Me_{N\text{-}S}$ = -0.10, $Me_{S\text{-}S}$ = -0.09). Kernel density estimate for S-S is omitted because there are only two observations for ETEMs.

The ETEMs differ also by types of measure (figure 4). A majority of papers focuses on measures aiming at protecting human health (17 papers on Sanitary and Phytosanitary Standards, SPSs, and 25 papers on Maximum Residue Levels, MRLs: 504 point estimates), while several studies (15 papers, for 362 point estimates) report evidence for NTMs not involved in specific categories (such as Technical Barriers to Trade, TBTs, SPSs, MRLs, or Specific Trade Concerns, STCs): these cases (grouped under the tag 'Other') show a large heterogeneity in estimates, ranging from more than fifty to less than minus fifteen (figure 4, panel (i)). The variabilities of ETEMs for papers that analyse different types of NTMs are similar ($\sigma_{TBT}$ = 3.51, $\sigma_{SPS}$ = 3.66, $\sigma_{Other}$ = 3.65), exception made for papers on MRLs, for which the variability is rather low ($\sigma_{MRL}$ = 2.20). While the median values of ETEMs associated with TBTs and MRLs are close to zero ($Me_{TBT}$ = 0.01, $Me_{MRL}$ = 0.02), the median for SPSs is negative ($Me_{SPS}$ = -0.11) (figure 4, panel (ii)).

Figure 4. Estimated trade effect of measures (ETEMs) arranged by types of measure.

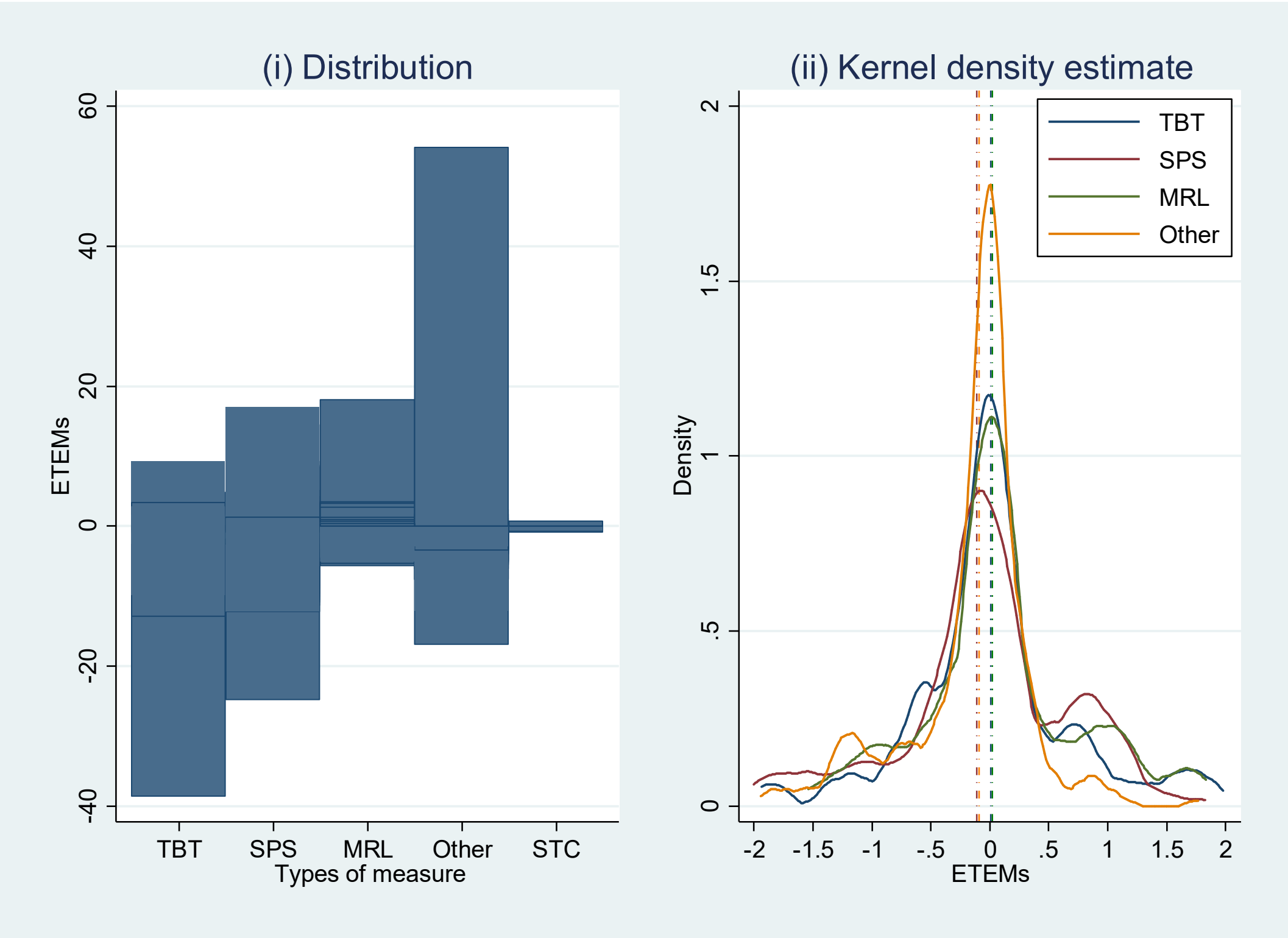

Notes: Types of measure are as follows: TBT stands for 'Technical Barrier to Trade', SPS stands for 'Sanitary and Phytosanitary Standard', MRL stands for 'Maximum Residue Level', STC stands for 'Specific Trade Concern', 'Other' includes measures not involved in other categories (e.g. quality and quantity control measures, Hazard Analysis and Critical Control Points (HACCP), private standards, voluntary standards). In panel (ii), the estimated densities for ETEMs are computed removing observations which exceed the $10^{th}$ and the $95^{th}$ percentiles. Dashed lines are median values (Me) computed on total observations (i.e. $Me_{Tot.}$ = -0.05, $Me_{TBT}$ = 0.01, $Me_{SPS}$ = -0.11, $Me_{MRL}$ = 0.02, $Me_{Other}$ = -0.09). Kernel density estimate for STC is omitted because there are only 24 observations for ETEMs.

### *4.2 Regression results*

We compare the results of robust regression, Multinomial Logit (MNL), Probit, and Tobit models The type of non-tariff measures (NTMs) does matter in determining magnitude, direction, and accuracy of the estimated trade effects of measures (ETEMs). If a study deepens on Technical Barriers to Trade (TBTs) or Sanitary and Phytosanitary Standards (SPSs), ETEMs tend to be greater, more accurate if significant, and significant negative with a lower probability (table 4, columns A, B, E, and F). ETEMs are greater also if Maximum Residue Level (MRL) is the measure under investigation, and are more likely to be significant positive, but less accurate, (table 4, columns A, C, and F). Our findings extend the evidence provided by Li and Beghin (2012): we disentangle the effects of TBTs and SPSs, and show the contribution of MRLs on trade.

As for the proxies used for NTMs, ETEMs tend to be lower and significant, regardless of the proxy (table 4, columns A and D). In particular, dummy variable and frequency index/coverage ratio are more likely to provide significant positive ETEMs (table 4, column B).

The level of detail of the study has a varying contribution on magnitude, significance, precision, and direction of ETEMs. Studies that deepen on NTMs across countries with similar levels of economic development ('N-N') tend to report ETEMs significantly different from zero and more accurate (table 4, columns D, E, and F). Our results expand the findings provided by Li and Beghin (2012, p. 507), who observe that "*when the NTM is a SPS policy regulating agri-food exports from a developing exporter to a developed importer, the probability to observe a trade impeding effect increases substantially*".

If data are aggregated at HS-2 digit or HS-4 digit, ETEMs are greater (table 4, column A): in particular, the higher the disaggregation, the higher the probability of significant ETEMs, that tend to be less accurate if significant negative (table 4, columns D and E).

As for specific product categories, ETEMs tend to be greater for cereal (table 4, column A), significant (either negative or positive) with higher probability for meat, but not for dairy (table 4, columns B, C, and D), more accurate if significant positive for fats and oils (table 4, columns F). In line with Li and Beghin (2012) we show that technical measures are not likely to be trade-enhancing for processed food products (e.g. dairy produce, fats and oils), while we also found that this is not always true (e.g. meat).

Table 4. Regression results.

| Covariates | Robust regression | | MNL | | | | Probit | | Tobit | | | |
|---|---|---|---|---|---|---|---|---|---|---|---|---|
| | $\hat{t}$ | | $\ln\left(\frac{Pr(Y_{MNL}=-1)}{Pr(Y_{MNL}=0)}\right)$ | | $\ln\left(\frac{Pr(Y_{MNL}=1)}{Pr(Y_{MNL}=0)}\right)$ | | $Pr\left(Y_{Probit}=1 \mid \mathrm{B}_{j,i}, \Gamma_{k,i}\right)$ | | $Y_{Tobit(-1.96)}$ | | $Y_{Tobit(1.96)}$ | |
| | A | | B | | C | | D | | E | | F | |
| $\beta_1$: TBT | 0.118 | *** | -0.180 | ** | 0.020 | | -0.016 | | 12.170 | * | 0.207 | |
| | (0.030) | | (0.070) | | (0.031) | | (0.018) | | (7.032) | | (0.250) | |
| $\beta_2$: SPS | 0355 | * | -0.036 | *** | 3.04 | | -4.450 | | 0.0002 | ** | 0.938 | ** |
| | (0.211) | | (0.334) | | (4.82) | | (3.070) | | (0.008) | | (0.389) | |
| $\beta_3$: MRL | 0.004 | *** | -0.024 | | 0.002 | ** | 0.001 | | 2.031 | | 0.020 | ** |
| | (0.002) | | (0.021) | | (0.001) | | (0.0004) | | (1.689) | | (0.008) | |
| $\beta_4$: AVE | -0.161 | ** | 0.130 | *** | 0.091 | * | 0.078 | *** | -1.837 | | 0.221 | |
| | (0.067) | | (0.045) | | (0.048) | | (0.024) | | (2.511) | | (0.335) | |
| $\beta_5$: Dummy variable | -0.130 | *** | 0.0619 | *** | 0.049 | | 0.039 | *** | -0.800 | | 0.333 | |
| | (0.018) | | (0.018) | | (0.036) | | (0.012) | | (1.633) | | (0.270) | |
| $\beta_6$: Index | -0.056 | *** | 0.033 | * | -0.002 | | 0.016 | * | -0.870 | | -0.069 | |
| | (0.011) | | (0.020) | | (0.026) | | (0.008) | | (0.566) | | (0.153) | |
| $\beta_7$: N-N | 0.004 | | -0.001 | | 0.009 | | 0.004 | * | 2.085 | *** | 0.116 | *** |
| | (0.015) | | (0.010) | | (0.010) | | (0.002) | | (0.218) | | (0.033) | |
| $\beta_8$: HS-2 digit | 0.098 | *** | -0.026 | | -0.042 | | -0.016 | | -0.280 | | -0.241 | |
| | (0.030) | | (0.042) | | (0.031) | | (0.014) | | (2.604) | | (0.303) | |
| $\beta_9$: HS-4 digit | 0.060 | *** | 0.007 | | 0.007 | | 0.002 | * | -1.904 | *** | 0.013 | |
| | (0.015) | | (0.008) | | (0.009) | | (0.001) | | (0.082) | | (0.017) | |
| $\beta_{10}$: Meat | 0.119 | | 0.397 | * | 0.469 | * | 0.275 | ** | -15.680 | | 1.342 | |
| | (0.139) | | (0.223) | | (0.265) | | (0.117) | | (11.730) | | (1.687) | |
| $\beta_{11}$: Dairy | -0.044 | | -0.183 | | -0.326 | | -0.147 | ** | 7.278 | | -3.143 | |
| | (0.088) | | (0.112) | | (0.238) | | (0.063) | | (8.311) | | (1.937) | |
| $\beta_{12}$: Cereal | -0.151 | * | -0.024 | | -0.295 | | -0.059 | | -0.221 | | -1.825 | |
| | (0.079) | | (0.079) | | (0.448) | | (0.042) | | (6.690) | | (2.039) | |
| $\beta_{13}$: F&O | -0.048 | | 0.059 | | 0.063 | | 0.031 | | -1.680 | | 2.991 | *** |
| | (0.042) | | (0.060) | | (0.045) | | (0.027) | | (1.671) | | (0.013) | |
| $\gamma_1$: Country-pair f.e. | -0.762 | ** | 1.561 | *** | 1.618 | ** | 0.948 | *** | -39.160 | | 8.520 | |
| | (0.311) | | (0.558) | | (0.813) | | (0.284) | | (43.530) | | (7.368) | |
| $\gamma_2$: Time f.e. | 0.786 | | 0.269 | | -0.113 | | 0.091 | | -3.658 | | -4.339 | |
| | (0.567) | | (0.523) | | (0.478) | | (0.277) | | (27.740) | | (4.674) | |
| $\gamma_3$: Product f.e. | -1.285 | *** | -0.008 | | -1.099 | * | -0.196 | | -16.010 | | -9.968 | * |
| | (0.239) | | (0.325) | | (0.631) | | (0.195) | | (22.260) | | (6.010) | |
| $\gamma_4$: Zero trade | -1.501 | *** | 0.672 | * | -0.019 | | 0.258 | | -38.140 | | -4.422 | |
| | (0.278) | | (0.404) | | (0.432) | | (0.182) | | (27.060) | | (4.836) | |
| $\gamma_5$: Q1 | -1.196 | *** | 0.070 | | -0.930 | ** | -0.209 | | -28.950 | | -11.640 | ** |
| | (0.357) | | (0.443) | | (0.408) | | (0.179) | | (25.440) | | (5.842) | |
| $\gamma_6$: WP | -0.718 | *** | -0.722 | | -1.469 | ** | -0.583 | | 13.930 | | -8.198 | |
| | (0.267) | | (0.576) | | (0.739) | | (0.358) | | (28.760) | | (7.864) | |
| $\gamma_7$: Authors | -0.856 | * | 1.335 | *** | 1.338 | * | 0.812 | *** | -50.060 | | 7.062 | |
| | (0.455) | | (0.447) | | (0.719) | | (0.221) | | (39.960) | | (6.401) | |
| Constant | 1.583 | *** | -1.816 | *** | -1.370 | * | -0.628 | *** | 176.400 | * | -14.790 | ** |
| | (0.298) | | (0.497) | | (0.758) | | (0.240) | | (95.610) | | (7.417) | |
| Sigma | | | | | | | | | 140.600 | ** | 19.020 | *** |
| | | | | | | | | | (70.580) | | (3.346) | |
| Observations | 1,210 | | 1,213 | | 1,213 | | 1,213 | | 1,213 | | 1,213 | |

Notes:

Clustered standard errors are in parentheses.

***, **, and * indicate statistical significance at 1%, 5%, and 10%.

The coefficient $\alpha_1$ has been omitted because of collinearity.

The magnitude of estimated coefficients and related standard errors for variables 'SPS' are of the order of $10^{-15}$.

Acronyms are as follows: North-North (N-N), North-South (N-S), Technical Barrier to Trade (TBT), Sanitary and Phytosanitary Standard (SPS), Maximum Residue Level (MRL), *ad valorem equivalent* (AVE), Harmonised System (HS), peer-reviewed journal ranked in the first quartile following the classification of Scimago Journal & Country Rank (SJR) at the date of publication and the subject area 'Economics and Econometrics'(Q1), working paper (WP).

Studies that include country-pair fixed effects provide lower and significant ETEMs (either negative or positive) with a higher probability (table 4, columns A, B, C, and D). Similarly, Li and Beghin (2012) pointed that the trade effects of technical measure are influenced by the use of multilateral trade resistance terms. ETEMs are lower also in studies with time fixed effects (table 4, column A), but significant positive with a lower probability and, in these cases, less accurate in studies with product fixed effects (table 4, columns C and F). On top of previous knowledge, we show that controlling for time and for product-specific (or sector/industry-specific) fixed effects impacts on magnitude, direction, and accuracy of ETEMs.

If a study accounts for the treatment of zero trade flows, ETEMs tend to be lower and the likelihood of negative ETEMs significantly different from zero increases (table 4, columns A and B). Similarly, Li and Beghin (2012, p. 507) argue that "*t-values are more spread out in the negative range when zero trade is treated*".

Reflecting on the publication process, ETEMs are lower and less likely to be positive significant if provided in studies published in top journals (Q1) or in working paper series (table 4, columns A and C). In particular, if significant positive, ETEMs are less accurate if published in Q1 (table 4, column F). In addition, ETEMs are lower and significantly different from zero (either positive or negative) with higher probability if a study is co-authored by experienced scholars (table 4, columns A, B, C, and D). Similarly, Havránek (2010, p. 254) argues that the authorship helps explaining the direction and the magnitude of estimates.

Our analysis deepens on several issues: a number of variables contribute to explain the heterogeneity in ETEMs. The magnitude of estimates is favoured by certain factors (type of NTMs, product aggregation), but limited by other determinants (proxy for NTMs, 'cereal', 'country-pair f.e.', 'time f.e.', 'zero trade', publication process). Some factors reduce the likelihood of having significant estimates ('dairy'), others intensify this likelihood (proxy for NTMs, 'authors', 'country-pair f.e.', 'N-N', 'HS-4 digit', 'meat'). Moreover, some variables boost (proxy for NTMs, 'authors', 'country-pair f.e.', 'zero trade', 'meat') and others hamper ('SPS', 'TBT') the probability of estimating trade-impeding effects. Similarly, the likelihood of estimating trade-enhancing effects may be either intensified ('authors', 'country-pair f.e.', 'MRL', 'AVE', 'meat') or limited ('Q1', 'WP', 'product f.e.') by specific variables. In addition, the accuracy of significant negative estimates increase with type of NTMs and decrease with 'HS-4 digit'; vice-versa, the accuracy of significant positive estimates is favoured by certain variables ('SPS', 'MRL', 'N-N', 'F&O'), but not by others ('product f.e.', 'Q1'). Table 5 synthesises the evidence of our empirical models.

Table 5. Summary of findings on the estimated trade effects of measures (ETEMs).

| Covariates | | Magnitude | Significance | Negative significance | Positive significance | Accuracy of negative significant ETEMs | Accuracy of positive significant ETEMs |
|---|---|---|---|---|---|---|---|
| Type of NTMs | TBT | Greater | n.s. | Less likely | n.s. | More accurate | n.s. |
| | SPS | Greater | n.s. | Less likely | n.s. | More accurate | More accurate |
| | MRL | Greater | n.s. | n.s. | More likely | More accurate | More accurate |
| Proxy for NTMs | AVE | Lower | More likely | More likely | More likely | n.s. | n.s. |
| | Dummy variable | Lower | More likely | More likely | n.s. | n.s. | n.s. |
| | Index | Lower | More likely | More likely | n.s. | n.s. | n.s. |
| Level of detail of the study | N-N | n.s. | More likely | n.s. | n.s. | n.s. | More accurate |
| | HS-2digit | Greater | n.s. | n.s. | n.s. | n.s. | n.s. |
| | HS-4digit | Greater | More likely | n.s. | n.s. | Less accurate | n.s. |
| | Meat | n.s. | More likely | More likely | More likely | n.s. | n.s. |
| | Dairy | n.s. | Less likely | n.s. | n.s. | n.s. | n.s. |
| | Cereal | Lower | n.s. | n.s. | n.s. | n.s. | n.s. |
| | F&O | n.s. | n.s. | n.s. | n.s. | n.s. | More accurate |
| Methodological issues | Country-pair f.e. | Lower | More likely | More likely | More likely | n.s. | n.s. |
| | Time f.e. | Lower | n.s. | n.s. | n.s. | n.s. | n.s. |
| | Product f.e. | n.s. | n.s. | n.s. | Less lilely | n.s. | Less accurate |
| | Zero trade | Lower | n.s. | More likely | n.s. | n.s. | n.s. |
| Publication process | Q1 | Lower | n.s. | n.s. | Less likely | n.s. | Less accurate |
| | WP | Lower | n.s. | n.s. | Less likely | n.s. | n.s. |
| | Authors | Lower | More likely | More likely | More likely | n.s. | n.s. |

Note: n.s. stands for ‘not significant’.

## 5. Conclusions and policy implications

The rapid growth of non-tariff measures (NTMs) has stimulated an interesting academic debate. Discriminating between the economics and the politics of NTMs is a challenge for academics and policymakers: theory suggests that NTMs may both stimulate and hinder trade (Swinnen, 2017). Accordingly, in literature, two opposite views prevail: ‘standards as barrier’ versus ‘standards as catalyst’, with the empirical evidence being quite heterogeneous.

In order to characterise the heterogeneity in estimates, we qualitatively and quantitatively reviewed the empirical literature on the effects of NTMs on global agri-food trade. We explain the differences in findings, in terms of magnitude, direction, statistical significance, and accuracy of estimates, with several control factors: types of NTMs, proxies used for NTMs, level of details of studies, methodological issues, and publication process. We build on the existing evidence provided in Li and Beghin (2012) with further details and focusing on the agri-food sector.

We found that the estimated trade effects of measures (ETEMs) are overestimated by types of NTMs, but underestimated by proxies of NTMs. Maximum Residue Levels (MRLs) and *ad valorem equivalent* (AVE) tend to favour trade, whereas it is not always true that Sanitary and Phytosanitary Standards (SPSs) and Technical Barriers to Trade (TBTs) limit trade.

The level of detail also matters: for instance, analysing trade between developed countries or working with disaggregated data plays in favour of significant estimates. Magnitude, significance, and accuracy of ETEMs may be also product-specific.

Last but not least, robust methodological approaches, and evidence provided by experienced authors, are correlated with higher chances of observing statistically significant, although underestimated ETEMs. Controls for specific methodological issues (e.g. inclusion of multilateral resistance terms, treatment of zero trade flows) is advisable.

Our results highlight that magnitude, direction, statistical significance, and accuracy of ETEMs are case-specific. This is in line with Livingston et al. (2008) who suggest that, in evaluating NTMs, economists try to compare benefits for trade and costs of management, production, market, and resource potentially related to an outbreak of disease or pest, finding difference on a case-by-case basis. Thus, the trade effects of NTMs are likely to depend on specific countries, products, and standards: generalisations are not feasible. A plausible explanation is that the variability in trade effects may reflect divergences among countries' food safety regulations and standards, differences in consumers' preferences across countries, ability (or limited capacity) to produce safe food, and willingness to pay for risk-reducing technology (Buzby and Mitchell, 2006; Jongwanich, 2009).

The present analysis helps distinguishing 'causes' and 'confounding factors' that hinder the effects of policy measures on trade, and it aims to support applied economists and policymakers in understanding how NTMs shape trade flows. Academics should consider the importance to use suitable proxy for NTMs and to control for specific methodological issues, in order to provide estimates the more reliable as possible. Policymakers should carefully evaluate the peculiar effects and consequences of introducing new NTMs or modifying the existent ones on trade in certain sectors or between certain countries.

## Appendix

### *A.1 A comparison with Li and Beghin (2012)*

Li and Beghin (2012) propose a meta-analysis to explain how the variation in estimated trade effects of technical barriers to trade is affected by a number of explanatory variables (e.g. data sampling, methodology differences). They analyse a set of 27 papers dealing with technical measures (i.e. Sanitary and Phytosanitary Standards, SPSs, Technical Barriers to Trade, TBTs, and Maximum Residue Levels, MRLs) and theoretically based on gravity model.

We analyse a newer, larger, and updated dataset, following the methodological arguments of meta-analysis (Stanley and Jarrell, 1989; Stanley, 2005; Stanley et al., 2008; Stanley et al., 2013). In fact, the number of studies published since 2012 has grown tremendously: in five years at least 49 papers have been published in peer-reviewed journals (see figure A.1). This growing trend calls for specific attention: are findings of Li and Beghin (2012) still valid or does empirical literature show a trend reversal?

Figure A.1. Trend of published researches on trade effects of NTMs over time.

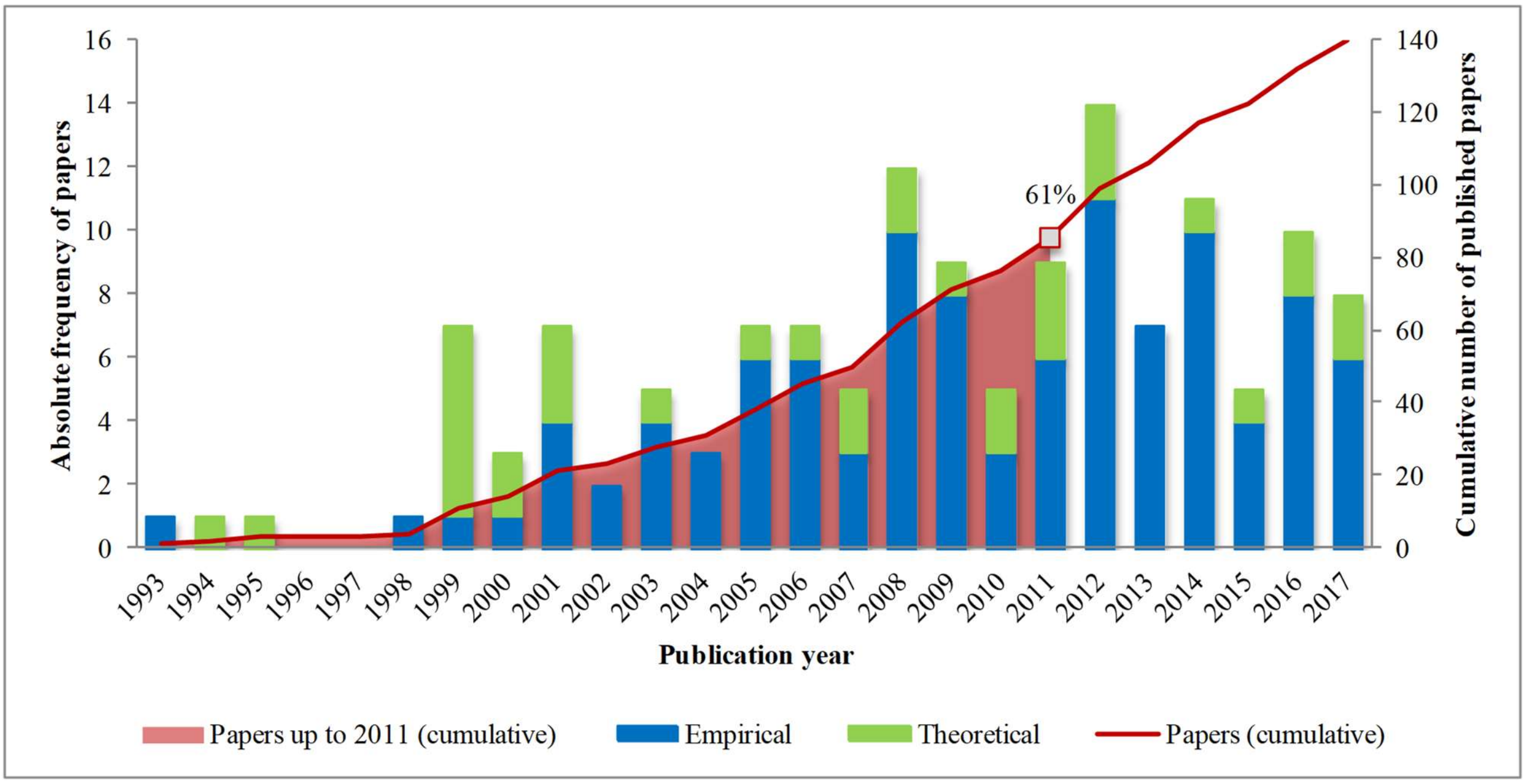

We analyse a wider sample in terms of number of papers considered (62 papers), type of measures investigated, and theoretical framework (not only gravity-based papers). As for the type of measures under investigation, in addition to SPSs, TBTs, MRLs, we include studies on standards that pursue similar scopes (e.g. protect consumers' health and safety, reducing asymmetric information): some examples are quality and quantity control measures, private standards, voluntary standards, requirements on Genetically Modified Organisms (GMOs).

We use a twice as large set of explanatory variables. We expand the number of variables able to influence the likelihood of acceptance of the study (7 vs. 2), and find that they matter. We also expand the number of variables that may affect magnitude and direction of the estimates (13 vs. 9). In particular, we use dummy variables to explain the influence of type of NTMs, proxy used for NTMs, and level of details of the study.

Following the same approach of Li and Beghin (2012), we adopt a robust regression technique and perform a Multinomial Logit (MNL) model. In addition, we present findings from a combination of Probit and Tobit models: the former allows us to disentangle the probability that a certain driver determines statistically significant estimates; the latter allows us to quantify the accuracy of significant estimates.

### *A.2 Additional notes to the selection of studies and exclusion criteria*

The search of relevant studies on 'trade and non-tariff measures (NTMs)' were performed through bibliographic databases (Scopus, Web of Science, JSTOR, RePEc, IATRC, AgEcon Search, Google Scholar) and repositories of field journals (table A.1).

Table A.1. List of field journals, classified according to their rank in Scimago Journal & Country Rank (SJR).

| Journal | SJR[a] |
|---|---|
| Agribusiness | Q2 |
| Agricultural Economics | Q1 |
| American Journal of Agricultural Economics | Q1 |
| Applied Economics Policy Perspective | Q2 |
| Australian Journal of Agricultural and Resource Economics | Q2 |
| Canadian Journal of Agricultural Economics | Q2 |
| China Agricultural Economic Review | Q3 |
| European Review of Agricultural Economics | Q1 |
| German Journal of Agricultural Economics | Q3 |
| Journal of Agricultural Economics | Q1 |
| Journal of Development Economics | Q1 |
| Journal of Development Studies[b] | Q1 |
| World Bank Economic Review | Q1 |
| World Development | Q1 |

[a] The rank is referred to the subject area 'Economics and Econometrics' at 2017.

[b] The rank is referred to the subject area 'Development' at 2017.

Table A.2 provides the list of the first 10 papers identified in Google Scholar on the basis of each keyword (i.e. 'agri-food trade' and 'non-tariff measure/non-tariff barrier'/'technical barrier to trade'/'sanitary and phytosanitary standard'/'maximum residue level'/'specific trade concern'). The same procedure has been used for each of bibliographic databases of references. The indicator, that ensures the inclusion of all relevant papers in the sample, is the frequency of appearance of a paper across the use of different keywords and sources of adoption.

Table A.2. Literature searching criteria adopted in each bibliographic database: example on Google Scholar.

| Keywords | | References |
|---|---|---|
| Agri-food trade | Non-tariff measures | 1. Disdier, A.C., & van Tongeren, F. (2010). Non-tariff measures in agri-food trade: What do the data tell us? Evidence from a cluster analysis on OECD imports. *Applied Economic Perspectives and Policy*, *32*(3), 436-455.<br>2. Chevassus-Lozza, E., Latouche, K., Majkovič, D., & Unguru, M. (2008). The importance of EU-15 borders for CEECs agri-food exports: The role of tariffs and non-tariff measures in the pre-accession period. *Food Policy*, *33*(6), 595-606.<br>3. Bureau, J.C., Marette, S., & Schiavina, A. (1998). Non-tariff trade barriers and consumers' information: The case of the EU-US trade dispute over beef. *European Review of Agricultural Economics*, *25*(4), 437-462.<br>4. Hooker, N.H., & Caswell, J.A. (1999). A Framework for Evaluating Non-Tariff Barriers to Trade Related to Sanitary and Phytosanitary Regulation. *Journal of Agricultural Economics*, *50*(2), 234-246.<br>5. Otsuki, T., Wilson, J.S., & Sewadeh, M. (2001). Saving two in a billion: quantifying the trade effect of European food safety standards on African exports. *Food policy*, *26*(5), 495-514.<br>6. Henson, S., & Loader, R. (2001). Barriers to agricultural exports from developing countries: the role of sanitary and phytosanitary requirements. *World Development*, *29*(1), 85-102.<br>7. Anderson, K., & Martin, W. (2005). Agricultural trade reform and the Doha Development Agenda. *The World Economy*, *28*(9), 1301-1327.<br>8. Disdier, A.C., Fontagné, L., & Mimouni, M. (2008). The impact of regulations on agricultural trade: evidence from the SPS and TBT agreements. *American Journal of Agricultural Economics*, *90*(2), 336-350.<br>9. Winchester, N., Rau, M.L., Goetz, C., Larue, B., Otsuki, T., Shutes, K., ... & Nunes de Faria, R. (2012). The impact of regulatory heterogeneity on agri-food trade. *The World Economy*, *35*(8), 973-993.<br>10. Beghin, J., Disdier, A.C., Marette, S., & Van Tongeren, F. (2012). Welfare costs and benefits of non-tariff measures in trade: a conceptual framework and application. *World Trade Review*, *11*(3), 356-375. |
| | Non-tariff barriers | 1. Bureau, J.C., Marette, S., & Schiavina, A. (1998). Non-tariff trade barriers and consumers' information: The case of the EU-US trade dispute over beef. *European Review of Agricultural Economics*, *25*(4), 437-462.<br>2. Hooker, N.H., & Caswell, J.A. (1999). A Framework for Evaluating Non-Tariff Barriers to Trade Related to Sanitary and Phytosanitary Regulation. *Journal of Agricultural Economics*, *50*(2), 234-246.<br>3. Henson, S., & Loader, R. (2001). Barriers to agricultural exports from developing countries: the role of sanitary and phytosanitary requirements. *World Development*, *29*(1), 85-102.<br>4. Otsuki, T., Wilson, J.S., & Sewadeh, M. (2001). Saving two in a billion:: quantifying the trade effect of European food safety standards on African exports. *Food policy*, *26*(5), 495-514.<br>5. Chevassus-Lozza, E., Latouche, K., Majkovič, D., & Unguru, M. (2008). The importance of EU-15 borders for CEECs agri-food exports: The role of tariffs and non-tariff measures in the pre-accession period. *Food Policy*, *33*(6), 595-606.<br>6. Disdier, A.C., Fontagné, L., & Mimouni, M. (2008). The impact of regulations on agricultural trade: evidence from the SPS and TBT agreements. *American Journal of Agricultural Economics*, *90*(2), 336-350.<br>7. Anderson, K., & Martin, W. (2005). Agricultural trade reform and the Doha Development Agenda. *The World Economy*, *28*(9), 1301-1327.<br>8. Lupien, J. R. (2002). The precautionary principle and other non-tariff barriers to free and fair international food trade. *Critical Reviews in Food Science and Nutrition*, *42*(4), 403-415.<br>9. Winchester, N. (2009). Is there a dirty little secret? Non-tariff barriers and the gains from trade. *Journal of Policy Modeling*, *31*(6), 819-834.<br>10. Henson, S., & Caswell, J. (1999). Food safety regulation: an overview of contemporary issues. *Food policy*, *24*(6), 589-603. |

Table A.2. (Continued).

| Keywords | | References |
|---|---|---|
| Agri-food trade | Technical Barriers to Trade | 1. Otsuki, T., Wilson, J.S., & Sewadeh, M. (2001). Saving two in a billion:: quantifying the trade effect of European food safety standards on African exports. *Food policy*, *26*(5), 495-514.<br>2. Li, Y., & Beghin, J. C. (2012). A meta-analysis of estimates of the impact of technical barriers to trade. *Journal of Policy Modeling*, *34*(3), 497-511.<br>3. Rose, A.K., & Van Wincoop, E. (2001). National money as a barrier to international trade: The real case for currency union. *American Economic Review*, *91*(2), 386-390.<br>4. Hobbs, J.E., & Kerr, W.A. (2006). Consumer information, labelling and international trade in agri-food products. *Food Policy*, *31*(1), 78-89.<br>5. Henson, S., & Loader, R. (2001). Barriers to agricultural exports from developing countries: the role of sanitary and phytosanitary requirements. *World development*, *29*(1), 85-102.<br>6. Beghin, J.C., & Bureau, J.C. (2001). Quantitative policy analysis of sanitary, phytosanitary and technical barriers to trade. *Économie internationale*, (3), 107-130.<br>7. Disdier, A.C., Fontagné, L., & Mimouni, M. (2008). The impact of regulations on agricultural trade: evidence from the SPS and TBT agreements. *American Journal of Agricultural Economics*, *90*(2), 336-350.<br>8. Roberts, M.B.D., Josling, T.E., & Orden, D. (1999). A framework for analyzing technical trade barriers in agricultural. *Technical Bulletin*, (1876).<br>9. Maertens, M., & Swinnen, J. (2006, January). Standards as barriers and catalysts for trade and poverty reduction. In *IAAE Conference Papers* (pp. 1-34).<br>10. Henson, S. (2008). The role of public and private standards in regulating international food markets. *Journal of International Agricultural Trade and Development*, *4*(1), 63-81. |
| | Sanitary and Phytosanitary Standards | 1. Henson, S., & Loader, R. (2001). Barriers to agricultural exports from developing countries: the role of sanitary and phytosanitary requirements. *World Development*, *29*(1), 85-102.<br>2. Otsuki, T., Wilson, J.S., & Sewadeh, M. (2001). Saving two in a billion: quantifying the trade effect of European food safety standards on African exports. *Food Policy*, *26*(5), 495-514.<br>3. Gebrehiwet, Y., Ngqangweni, S., & Kirsten, J.F. (2007). Quantifying the trade effect of sanitary and phytosanitary regulations of OECD countries on South African food exports. *Agrekon*, *46*(1), 1-17.<br>4. Henson, S., & Loader, R. (1999). Impact of sanitary and phytosanitary standards on developing countries and the role of the SPS Agreement. *Agribusiness*, *15*(3), 355-369.<br>5. Caswell, J.A., & Hooker, N.H. (1996). HACCP as an international trade standard. *American Journal of Agricultural Economics*, *78*(3), 775-779.<br>6. Disdier, A.C., Fontagné, L., & Mimouni, M. (2008). The impact of regulations on agricultural trade: evidence from the SPS and TBT agreements. *American Journal of Agricultural Economics*, *90*(2), 336-350.<br>7. Hooker, N.H., & Caswell, J.A. (1999). A Framework for Evaluating Non-Tariff Barriers to Trade Related to Sanitary and Phytosanitary Regulation. *Journal of Agricultural Economics*, *50*(2), 234-246.<br>8. Josling, T., Roberts, D., & Orden, D. (2004, August). Food regulation and trade: toward a safe and open global system-an overview and synopsis. In *American Agricultural Economics Association Annual Meeting*.<br>9. Henson, S. (2008). The role of public and private standards in regulating international food markets. *Journal of International Agricultural Trade and Development*, *4*(1), 63-81.<br>10. Unnevehr, L.J. (2000). Food safety issues and fresh food product exports from LDCs. *Agricultural Economics*, *23*(3), 231-240. |

Table A.2. (Continued).

| Keywords | | References |
|---|---|---|
| Agri-food trade | Maximum Residue Levels | 1. Winchester, N., Rau, M.L., Goetz, C., Larue, B., Otsuki, T., Shutes, K., ... & Nunes de Faria, R. (2012). The impact of regulatory heterogeneity on agri-food trade. *The World Economy*, *35*(8), 973-993.<br>2. Wilson, J.S., & Otsuki, T. (2004). To spray or not to spray: pesticides, banana exports, and food safety. *Food Policy*, *29*(2), 131-146.<br>3. Li, Y., & Beghin, J. C. (2014). Protectionism indices for non-tariff measures: An application to maximum residue levels. *Food Policy*, *45*, 57-68.<br>4. Wilson, J.S., Otsuki, T., & Majumdsar, B. (2003). Balancing food safety and risk: do drug residue limits affect international trade in beef?. *Journal of International Trade & Economic Development*,v*12*(4), 377-402.<br>5. Roitner-Schobesberger, B., Darnhofer, I., Somsook, S., & Vogl, C. R. (2008). Consumer perceptions of organic foods in Bangkok, Thailand. *Food policy*, *33*(2), 112-121.<br>6. Trienekens, J., & Zuurbier, P. (2008). Quality and safety standards in the food industry, developments and challenges. *International Journal of Production Economics*, *113*(1), 107-122.<br>7. Xiong, B., & Beghin, J. (2014). Disentangling demand-enhancing and trade-cost effects of maximum residue regulations. *Economic Inquiry*, *52*(3), 1190-1203.<br>8. Drogué, S., & DeMaria, F. (2012). Pesticide residues and trade, the apple of discord?. *Food Policy*, *37*(6), 641-649.<br>9. Jaffee, S.M., & Henson, S. (2005). Agro-food exports from developing countries: the challenges posed by standards. *Global Agricultural Trade and Developing Countries*, 91-114.<br>10. Henson, S., & Humphrey, J. (2010). Understanding the complexities of private standards in global agri-food chains as they impact developing countries. *The Journal of Development Studies*, *46*(9), 1628-1646. |
| | Specific Trade Concerns | 1. Caswell, J.A., & Hooker, N.H. (1996). HACCP as an international trade standard. *American Journal of Agricultural Economics*, *78*(3), 775-779.<br>2. Winchester, N., Rau, M.L., Goetz, C., Larue, B., Otsuki, T., Shutes, K., ... & Nunes de Faria, R. (2012). The impact of regulatory heterogeneity on agri-food trade. *The World Economy*, *35*(8), 973-993.<br>3. Vatn, A. (2002). Multifunctional agriculture: some consequences for international trade regimes. *European Review of Agricultural Economics*, *29*(3), 309-327.<br>4. Disdier, A.C., & van Tongeren, F. (2010). Non-tariff measures in agri-food trade: What do the data tell us? Evidence from a cluster analysis on OECD imports. *Applied Economic Perspectives and Policy*, *32*(3), 436-455.<br>5. Hoekman, B., & Anderson, K. (2000). Developing-country agriculture and the new trade agenda. *Economic Development and Cultural Change*, *49*(1), 171-180.<br>6. Henson, S., & Humphrey, J. (2010). Understanding the complexities of private standards in global agri-food chains as they impact developing countries. *The Journal of Development Studies*, *46*(9), 1628-1646.<br>7. Otsuki, T., Wilson, J.S., & Sewadeh, M. (2001). Saving two in a billion: quantifying the trade effect of European food safety standards on African exports. *Food Policy*, *26*(5), 495-514.<br>8. Josling, T. (2006). The war on terroir: geographical indications as a transatlantic trade conflict. *Journal of Agricultural Economics*, *57*(3), 337-363.<br>9. Hobbs, J.E. (2010). Public and private standards for food safety and quality: international trade implications. *The Estey Centre Journal of International Law and Trade Policy*, *11*(1), 136.<br>10. Jayasinghe, S., & Sarker, R. (2008). Effects of regional trade agreements on trade in agrifood products: Evidence from gravity modeling using disaggregated data. *Review of Agricultural Economics*, *30*(1), 61-81. |

Relevant studies were selected on the basis of information available in titles, abstracts, and full texts (figure A.2). 62 empirical researches (peer-reviewed papers, 70%, and gray literature, 30%) meet the inclusion criteria (figure A.2). We included empirical researches that assess the trade effects of NTMs, while we excluded theoretical papers, as well as papers that provide not comparable empirical results (table A.3).

Figure A.2. Literature searching criteria.

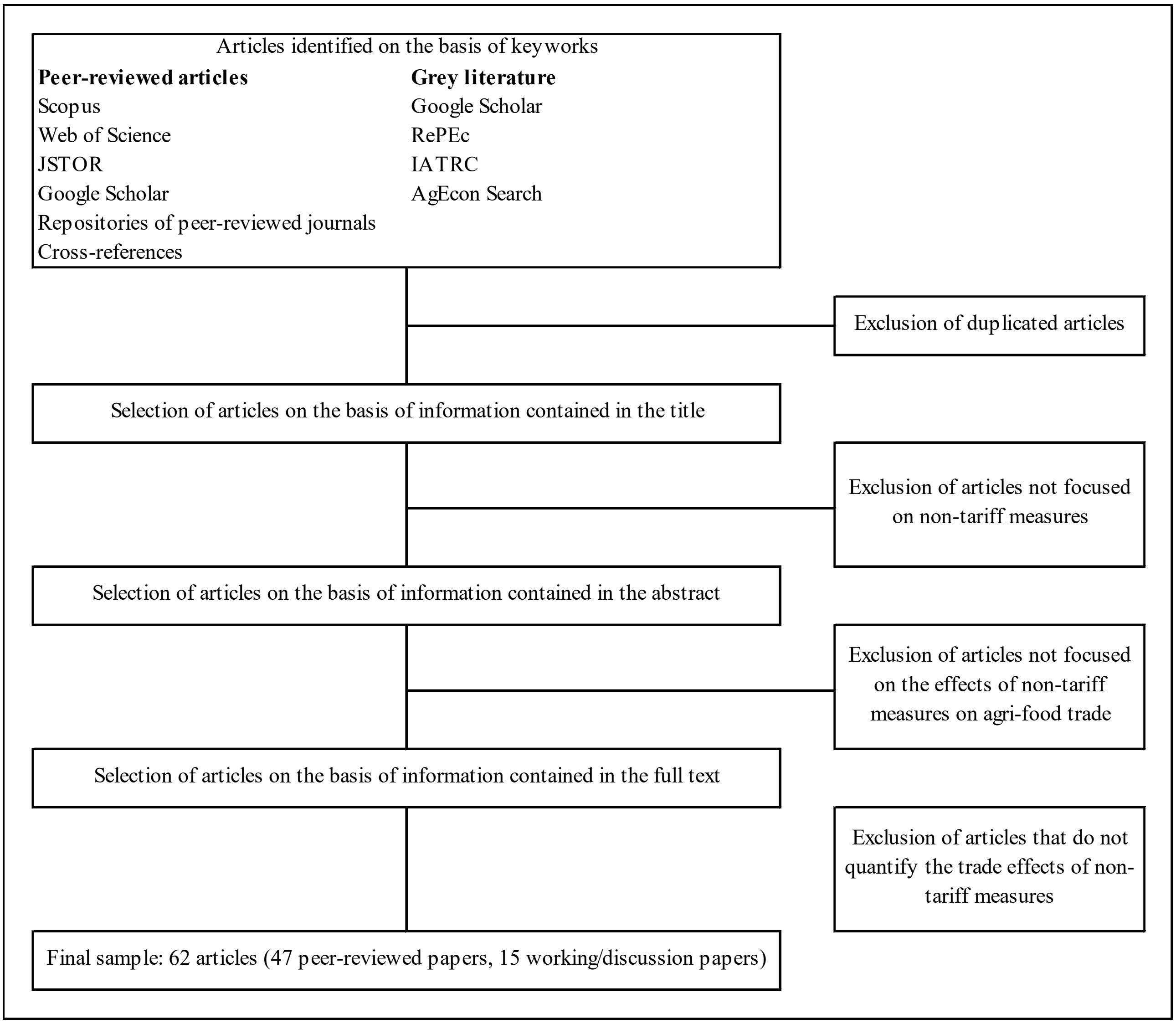

Figure A.3. Trend of empirical researches on trade effects of Non-Tariff Measures (NTMs) over time.

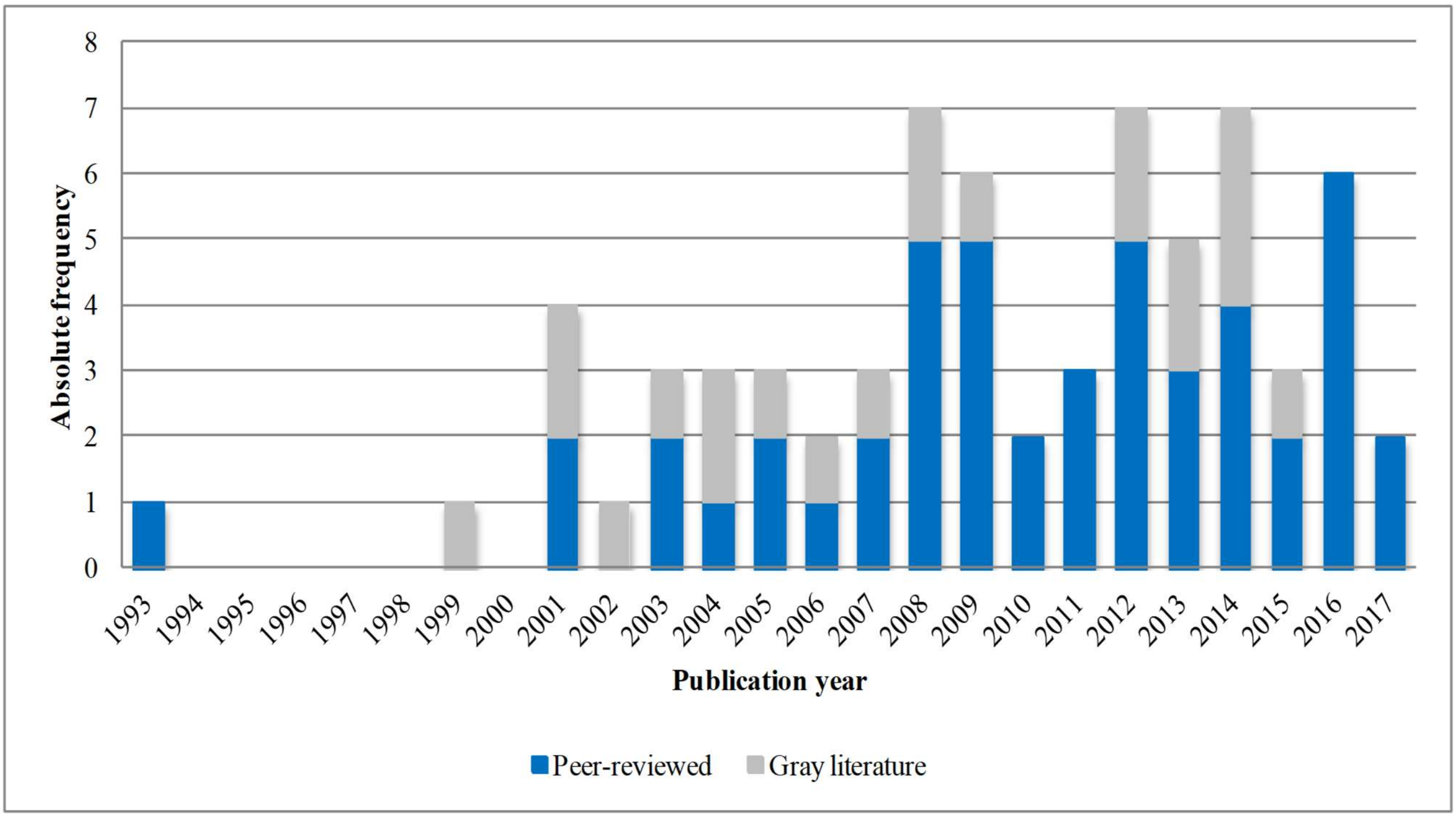

Table A.3. List of papers excluded from the meta-analysis.

| | |
|---|---|
| **Theoretical and/or descriptive papers on NTMs** | |
| 1 | Amjadi, A. and Yeats, A. (1995). Non-tariff Barriers Facing Africa: The Uruguay Round. *South African Journal of Economics 63*(3): 212-224. |
| 2 | Arita, S., Beckman, J., Kuberka, L. and Melton, A. (2014). Sanitary and phytosanitary measures and tariff-rate quotas for US meat exports to the European Union. Outlook Report No. LDPM-245-01. US Department of Agriculture, Economic Research Service. www.ers.usda.gov/publications/ldpm-livestock,-dairy,-and-poultry-outlook/ldpm-245-01.aspx. |
| 3 | Athukorala, P.C. and Jayasuriya, S. (2003). Food safety issues, trade and WTO rules: a developing country perspective. *The World Economy 26*(9): 1395-1416. |
| 4 | Bureau, J.C. and Doussin, J.P. (1999). Sanitary and technical regulations: Issues for trade liberalization in the dairy sector. *Canadian Journal of Agricultural Economics 47*(5): 149-156. |
| 5 | Clark, D.P. and Zarrilli, S. (1994). Non-tariff measures and United States' imports of CBERA-eligible products. *Journal of Development Studies* 31(1): 214-224. |
| 6 | de Almeida, F.M., da Cruz Vieira, W. and da Silva, O.M. (2012). SPS and TBT agreements and international agricultural trade: retaliation or cooperation?. *Agricultural Economics 43*(2): 125-132. |
| 7 | Donovan, J.A., Caswell, J.A. and Salay, E. (2001). The effect of stricter foreign regulations on food safety levels in developing countries: a study of Brazil. *Review of Agricultural Economics 23*(1): 163-175. |
| 8 | Edwards, G. and Fraser, I. (2001). Reconsidering agri-environmental policy permitted by the Uruguay round agreement. *Ecological Economics 37*(2): 313-326. |
| 9 | Fontagné, L., Von Kirchbach, F. and Mimouni, M. (2005). An Assessment of Environmentally-related Non-tariff Measures. *The World Economy 28*(10): 1417-1439. |
| 10 | Henson, S. and Humphrey, J. (2010). Understanding the complexities of private standards in global agri-food chains as they impact developing countries. *The Journal of Development Studies 46*(9): 1628-1646. |

Table A.3. (Continued).

| **Theoretical and/or descriptive papers on NTMs** | |
|---|---|
| 11 | Henson, S. and Jaffee, S. (2006). Food safety standards and trade: Enhancing competitiveness and avoiding exclusion of developing countries. *The European Journal of Development Research 18*(4): 593-621. |
| 12 | Henson, S. and Jaffee, S. (2008). Understanding developing country strategic responses to the enhancement of food safety standards. *The World Economy 31*(4): 548-568. |
| 13 | Henson, S. and Loader, R. (1999). Impact of sanitary and phytosanitary standards on developing countries and the role of the SPS Agreement. *Agribusiness 15*(3): 355-369. |
| 14 | Henson, S. and Loader, R. (2001). Barriers to agricultural exports from developing countries: the role of sanitary and phytosanitary requirements. *World Development 29*(1): 85-102. |
| 15 | Henson, S., Brouder, A.M. and Mitullah, W. (2000). Food safety requirements and food exports from developing countries: the case of fish exports from Kenya to the European Union. *American Journal of Agricultural Economics 82*(5): 1159-1169. |
| 16 | Henson, S., Masakure, O. and Cranfield, J. (2011). Do fresh produce exporters in sub-Saharan Africa benefit from GlobalGAP certification?. *World Development 39*(3): 375-386. |
| 17 | Hooker, N.H. (1999). Food safety regulation and trade in food products. *Food Policy 24*(6): 653-668. |
| 18 | Hooker, N.H. and Caswell, J.A. (1999). A Framework for Evaluating Non-Tariff Barriers to Trade Related to Sanitary and Phytosanitary Regulation. *Journal of Agricultural Economics 50*(2): 234-246. |
| 19 | Jouanjean, M.A., Maur, J.C. and Shepherd, B. (2016). US phytosanitary restrictions: the forgotten non-tariff barrier. *Journal of International Trade Law and Policy 15*(1): 2-27. |
| 20 | Kalaba, M., Sacolo, T. and Kirsten, J. (2016). Non-tariff measures affecting agricultural trade in SADC. *Agrekon 55*(4): 377-410. |
| 21 | Kareem, O.I. (2011). The European Union Trade Policies and Africa's Exports. *World Economics 12*(2): 54. |
| 22 | Koch, S.F. and Peet, M.A. (2007). Non-tariff barriers faced by South African firms: are there any lessons?: management. *South African Journal of Economic and Management Sciences 10*(4): 530-543. |
| 23 | Orden, D. and Roberts, D. (2007). Food regulation and trade under the WTO: ten years in perspective. *Agricultural Economics 37*(1): 103-118. |
| 24 | Rau, M.L. and Schlueter, S.W. (2009, August). Framework for analyzing regulations and standards in the NTM-Impact project. In International Association of Agricultural Economists Conference. IAAE. Beijing, China. |
| 25 | Rickard, B.J. and Lei, L. (2011). How important are tariffs and nontariff barriers in international markets for fresh fruit?. *Agricultural Economics 42*(1): 19-32. |
| 26 | Roberts, D. (1999). Analyzing technical trade barriers in agricultural markets: challenges and priorities. *Agribusiness 15*(3): 335-354. |
| 27 | Roberts, M.B.D., Josling, T.E. and Orden, D. (1999). A framework for analyzing technical trade barriers in agricultural. Technical Bulletin No. 1876. |
| 28 | Salvatici, L. and Nenci, S. (2017). New features, forgotten costs and counterfactual gains of the international trading system. *European Review of Agricultural Economics 44*(4): 592-633. |
| 29 | Staiger, R.W. (2012). Non-tariff measures and the WTO. Working Paper ERSD-2012-01. |
| 30 | Weyerbrock, S. and Xia, T. (2000). Technical trade barriers in US/Europe agricultural trade. *Agribusiness 16*(2): 235-251. |
| **Empirical papers on NTMs excluded from the regression analysis** | |
| 31 | Blind, K. (2001). The impacts of innovations and standards on trade of measurement and testing products: Empirical results of Switzerland's bilateral flows with Germany, France, and the UK. *Information Economics and Policy 13*: 439–460. |
| 32 | Blind, K. and Jungmittag, A. (2005). Trade and the impact of innovations and standards: The case of Germany and the UK. *Applied Economics 37*: 1385–1398. |
| 33 | Boere, E., Peerlings, J., Reinhard, S. and Heijman, W. (2015). The dynamics of dairy land use change with respect to the milk quota regime. European Review of Agricultural Economics 42(4): 651-674. |

Table A.3. (Continued).

| Empirical papers on NTMs excluded from the regression analysis | |
|---|---|
| 34 | Çakır, M., Boland, M.A. and Wang, Y. (2017). The Economic Impacts of 2015 Avian Influenza Outbreak on the US Turkey Industry and the Loss Mitigating Role of Free Trade Agreements. *Applied Economic Perspectives and Policy* [in press]. |
| 35 | Calvin, L. and Krissoff, B. (1998). Technical barriers to trade: a case study of phytosanitary barriers and US-Japanese apple trade. *Journal of Agricultural and Resource Economics 23*(2): 351-366. |
| 36 | Calvin, L., Krissoff, B. and Foster, W. (2008). Measuring the costs and trade effects of phytosanitary protocols: A US–Japanese apple example. *Review of Agricultural Economics 30*(1): 120-135. |
| 37 | Cioffi, A., Santeramo, F.G. and Vitale, C.D. (2011). The price stabilization effects of the EU entry price scheme for fruit and vegetables. *Agricultural Economics 42*(3): 405-418. |
| 38 | Cook, D.C., Carrasco, L.R., Paini, D.R. and Fraser, R.W. (2011). Estimating the social welfare effects of New Zealand apple imports. *Australian Journal of Agricultural and Resource Economics 55*(4): 599-620. |
| 39 | Crivelli, P. and Gröschl, J. (2012). *SPS measures and trade: Implementation matters* WTO Staff Working Paper No. ERSD-2012-05. |
| 40 | de Camargo Barros, G.S., Burnquist, H.L., de Miranda, S.H.G. and da Cunha Filho, J.H. (2002). SPS in Agricultural Trade: Issues and Options for a Research Agenda. |
| 41 | Disdier, A.C. and van Tongeren, F. (2010). Non-tariff measures in agri-food trade: What do the data tell us? Evidence from a cluster analysis on OECD imports. *Applied Economic Perspectives and Policy 32*(3): 436-455. |
| 42 | Disdier, A.C., Fontagné, L. and Cadot, O. (2014). North-South standards harmonization and international trade. *The World Bank Economic Review 29*(2): 327-352. |
| 43 | Engler, A., Nahuelhual, L., Cofré, G. and Barrena, J. (2012). How far from harmonization are sanitary, phytosanitary and quality-related standards? An exporter's perception approach. *Food Policy 37*(2): 162-170. |
| 44 | Fontagné, L., Mimouni, M. and Pasteels, J. M. (2005). Estimating the impact of environmental SPS and TBT on international trade. *Integration and Trade 22*: 7–37. |
| 45 | Fugazza, M. and Maur, J.C. (2008). Non-tariff barriers in CGE models: How useful for policy?. *Journal of Policy Modeling 30*(3): 475-490. |
| 46 | Grant, J.H. and Meilke, K.D. (2006). The World Trade Organization special safeguard mechanism: a case study of wheat. *Applied Economic Perspectives and Policy 28*(1): 24-47. |
| 47 | Hallak, J.C. (2006). Product quality and the direction of trade. *Journal of international Economics 68*(1): 238-265. |
| 48 | Haveman, J. and Thursby, J.G. (1999). The impact of tariff and non-tariff barriers to trade in agricultural commodities: a disaggregated approach. Purdue University 1-21. |
| 49 | Haveman, J.D., Nair-Reichert, U. and Thursby, J.G. (2003). How effective are trade barriers? An empirical analysis of trade reduction, diversion, and compression. *The Review of Economics and Statistics 85*(2): 480-485. |
| 50 | Jaud, M., Cadot, O. and Suwa-Eisenmann, A. (2013). Do food scares explain supplier concentration? An analysis of EU agri-food imports. *European Review of Agricultural Economics 40*(5): 873-890. |
| 51 | Junker, F. and Heckelei, T. (2012). TRQ-complications: who gets the benefits when the EU liberalizes Mercosur's access to the beef markets?. *Agricultural Economics 43*(2): 215-231. |
| 52 | Li, Y. and Beghin, J.C. (2014). Protectionism indices for non-tariff measures: An application to maximum residue levels. *Food Policy 45*: 57-68. |
| 53 | Maertens, M. and Swinnen, J.F. (2009). Trade, standards, and poverty: Evidence from Senegal. *World development 37*(1): 161-178. |
| 54 | Masakure, O., Henson, S. and Cranfield, J. (2009). Standards and export performance in developing countries: Evidence from Pakistan. *The Journal of International Trade & Economic Development 18*(3): 395-419. |
| 55 | Maskus, K.E., Wilson, J.S. and Otsuki, T. (2000). Quantifying the impact of technical barriers to trade. World Bank Policy Research Working Paper 2512. |

Table A.3. (Continued).

| | |
|---|---|
| **Empirical papers on NTMs excluded from the regression analysis** | |
| 56 | Nimenya, N., Ndimira, P.F. and de Frahan, B.H. (2012). Tariff equivalents of nontariff measures: the case of European horticultural and fish imports from African countries. *Agricultural Economics 43*(6): 635-653. |
| 57 | Orefice, G. (2017). Non-Tariff Measures, Specific Trade Concerns and Tariff Reduction. *The World Economy 40*(9): 1807-1835. |
| 58 | Otsuki, T., Wilson, J.S. and Sewadeh, M. (2001). A race to the top? a case study of food safety standards and African exports. The World Bank Working Paper. |
| 59 | Péridy, N. (2005). Towards a New Trade Policy Between the USA and Middle-East Countries: Estimating Trade Resistance and Export Potential. *The World Economy 28*(4): 491-518. |
| 60 | Rafajlovic, J. and Cardwell, R. (2013). The effects of a new WTO agreement on Canada's chicken market: A differentiated products modeling approach. *Canadian Journal of Agricultural Economics 61*(4): 487-507. |
| 61 | Raimondi, V. and Olper, A. (2011). Trade elasticity, gravity and trade liberalisation: evidence from the food industry. *Journal of Agricultural Economics 62*(3): 525-550. |
| 62 | Rial, D.P. (2014). Study of Average Effects of Non-tariff Measures on Trade Imports. UNCTAD. |
| 63 | Rickard, B.J. and Sumner, D.A. (2008). Domestic support and border measures for processed horticultural products. *American Journal of Agricultural Economics 90*(1): 55-68. |
| 64 | Santeramo, F.G. and Cioffi, A. (2012). The entry price threshold in EU agriculture: Deterrent or barrier?. *Journal of Policy Modeling 34*(5): 691-704. |
| 65 | Schuster, M. and Maertens, M. (2012). Do Private Standards Create Exclusive Food Supply Chains? New Evidence from the Peruvian Asparagus Export Sector. International Agricultural Trade Research Consortium No. 142769. |
| 66 | Soon, B.M. and Thompson, W. (2016). Measuring Non-Tariff Barriers by Combining Cointegration Tests and Simulation Models with an Application to Russian Chicken Imports. In Agricultural and Applied Economics Association Annual Meeting, Boston, Massachusetts No. 235738. |
| 67 | Swinnen, J. (2016). Economics and politics of food standards, trade, and development. *Agricultural Economics 47*(1): 7-19. |
| 68 | Swinnen, J. (2017). Some dynamic aspects of food standards. *American Journal of Agricultural Economics 99*(2): 321-338. |
| 69 | Taghouti, I., Alvarez-Coque, J.M.G. and Martinez-Gomez, V. (2017). Implications of changing aflatoxin standards for EU border controls on nut imports. In XV EAAE Congress, Parma, Italy. |
| 70 | van Veen, T.W.S. (2005). International trade and food safety in developing countries. *Food Control 16*(6): 491-496. |
| 71 | Wilson, J.S., and Otsuki, T. (2002). To Spray or Not to Spray: Pesticides, Banana Exports, and Food Safety. World Bank Working Paper No. 2805. |
| 72 | Wossink, A. and Gardebroek, C. (2006). Environmental policy uncertainty and marketable permit systems: The Dutch phosphate quota program. *American Journal of Agricultural Economics 88*(1): 16-27. |
| 73 | Yue, C., Beghin, J. and Jensen, H.H. (2006). Tariff equivalent of technical barriers to trade with imperfect substitution and trade costs. *American Journal of Agricultural Economics 88*(4): 947-960. |
| **Empirical papers not directly related to NTMs excluded from the regression analysis** | |
| 74 | Bertola, G. and Faini, R. (1990). Import demand and non-tariff barriers: the impact of trade liberalization: an application to Morocco. *Journal of Development Economics 34*(1-2): 269-286. |
| 75 | Boulanger, P., Dudu, H., Ferrari, E. and Philippidis, G. (2016). Russian Roulette at the Trade Table: A Specific Factors CGE Analysis of an Agri-food Import Ban. *Journal of Agricultural Economics 67*(2): 272-291. |
| 76 | Chen, N. (2004). Intra-national versus international trade in the European Union: Why do national borders matter? *Journal of International Economics 63*: 93–118. |
| 77 | Díaz-Bonilla, E., Robinson, S. and Swinnen, J.F. (2003). Regional agreements and the World Trade Organization negotiations. *American Journal of Agricultural Economics 85*(3): 679-683. |

Table A.3. (Continued).

## *A.3 Description of the sample*

Table A.4. Descriptive statistics for papers included in the sample.

| References | Obs. | Positive obs. | Positive significant obs. | Negative obs. | Negative significant obs. | Me | μ | σ | Min | Max |
|---|---|---|---|---|---|---|---|---|---|---|
| Anders and Caswell (2009) | 17 | 3 | 1 | 14 | 12 | -0.42 | -0.35 | 0.40 | -0.92 | 0.50 |
| Arita et al. (2017) | 10 | - | - | 10 | 8 | -1.00 | -1.78 | 1.77 | -5.49 | -0.53 |
| Babool and Reed (2007) | 1 | 1 | 1 | - | - | 0.98 | 0.98 | - | 0.98 | 0.98 |
| Beckman and Arita (2016) | 15 | 1 | - | 14 | 13 | -3.83 | -4.73 | 4.57 | -16.90 | 2.01 |
| Cardamone (2011) | 4 | 4 | 2 | - | - | 1.76 | 5.36 | 8.04 | 0.54 | 17.36 |
| Chen et al. (2008) | 5 | 5 | 5 | - | - | 0.28 | 0.42 | 0.33 | 0.21 | 1.00 |
| Chevassus-Lozza et al. (2008) | 58 | 7 | 6 | 51 | 50 | -0.38 | -0.38 | 0.39 | -1.32 | 0.24 |
| Crivelli and Gröschl (2016) | 72 | 46 | 34 | 26 | 11 | 0.07 | 0.27 | 0.52 | -0.58 | 1.00 |
| Dal Bianco et al. (2015) | 35 | 5 | - | 30 | 12 | -0.16 | -0.25 | 0.37 | -1.34 | 0.50 |
| de Frahan and Vancauteren (2006) | 11 | 10 | 10 | 1 | 1 | 1.55 | 1.48 | 1.04 | -0.33 | 3.43 |
| Disdier and Fontagné (2008) | 65 | 6 | - | 59 | 46 | -2.30 | -2.50 | 2.03 | -7.59 | 1.03 |
| Disdier and Marette (2010) | 3 | 3 | 3 | - | - | 0.15 | 0.14 | 0.01 | 0.13 | 0.15 |
| Disdier et al. (2008a) | 80 | 22 | 4 | 58 | 20 | -0.25 | -0.14 | 0.91 | -1.91 | 5.11 |
| Disdier et al. (2008b) | 64 | 33 | 17 | 31 | 22 | 0.01 | 0.02 | 1.12 | -3.19 | 3.15 |
| Drogué and DeMaria (2012) | 8 | 3 | - | 5 | 3 | -0.04 | -0.04 | 0.12 | -0.23 | 0.12 |
| Essaji (2008) | 2 | - | - | 2 | 2 | -0.05 | -0.05 | 0.00 | -0.05 | -0.05 |
| Fernandes et al. (2017) | 5 | - | - | 5 | 4 | -0.17 | -0.21 | 0.10 | -0.38 | -0.16 |
| Ferro et al. (2013) | 15 | 2 | 1 | 13 | 5 | -0.96 | -1.03 | 2.11 | -5.68 | 4.46 |
| Ferro et al. (2015) | 10 | 6 | 1 | 4 | 3 | 0.14 | 1.11 | 2.88 | -0.23 | 9.22 |
| Fontagné et al. (2005) | 8 | 1 | - | 7 | 6 | -0.57 | -0.61 | 0.81 | -2.13 | 0.85 |
| Gebrehiwet (2007) | 2 | 2 | 2 | - | - | 0.39 | 0.39 | 0.03 | 0.37 | 0.41 |
| Harrigan (1993) | 10 | - | - | 10 | 10 | -1.52 | -1.89 | 1.04 | -3.79 | -0.72 |
| Hoekman and Nicita (2011) | 11 | - | - | 11 | 10 | -0.17 | -0.13 | 0.06 | -0.19 | -0.03 |
| Jayasinghe et al. (2010) | 9 | 9 | 8 | - | - | 0.35 | 0.39 | 0.06 | 0.34 | 0.48 |
| Jongwanich (2009) | 1 | 1 | 1 | - | - | 0.05 | 0.05 | - | 0.05 | 0.05 |
| Kareem (2014a) | 2 | 1 | 1 | 1 | 1 | 0.81 | 0.81 | 3.55 | -1.71 | 3.32 |

Table A.4. Descriptive statistics for papers included in the sample.

| References | Obs. | Positive obs. | Positive significant obs. | Negative obs. | Negative significant obs. | Me | μ | σ | Min | Max |
|---|---|---|---|---|---|---|---|---|---|---|
| Kareem (2014b) | 4 | 1 | 1 | 3 | 2 | -1.59 | -3.16 | 6.42 | -12.16 | 2.71 |
| Kareem (2014c) | 4 | 1 | 1 | 3 | 2 | -1.59 | -3.16 | 6.42 | -12.16 | 2.71 |
| Kareem (2016a) | 8 | 5 | 1 | 3 | 2 | 0.00 | -0.42 | 0.99 | -2.85 | 0.07 |
| Kareem (2016b) | 2 | 1 | 1 | 1 | 1 | 1.76 | 1.76 | 6.11 | -2.57 | 6.08 |
| Kareem (2016c) | 5 | 4 | | 1 | - | 0.00 | 10.27 | 24.55 | -2.85 | 54.14 |
| Kareem et al. (2015) | 12 | 4 | 3 | 8 | 7 | -0.07 | 3.72 | 6.91 | -1.51 | 18.11 |
| Melo et al. (2014) | 5 | 1 | 1 | 4 | 2 | -0.41 | -0.14 | 0.52 | -0.46 | 0.76 |
| Moenius (2004) | 80 | 56 | 56 | 24 | 24 | 0.06 | 0.05 | 0.28 | -1.29 | 0.64 |
| Moenius (2006) | 33 | 18 | 18 | 15 | 15 | 0.09 | 0.16 | 0.42 | -0.49 | 1.32 |
| Munasib and Roy (2013) | 8 | 7 | - | 1 | - | 0.01 | 0.05 | 0.09 | -0.09 | 0.19 |
| Nardella and Boccaletti (2003) | 37 | 14 | 5 | 23 | 12 | -0.31 | -0.55 | 1.65 | -5.12 | 2.84 |
| Nardella and Boccaletti (2004) | 62 | 16 | 11 | 46 | 44 | -2.56 | -2.52 | 6.44 | -24.82 | 16.97 |
| Nardella and Boccaletti (2005) | 90 | 48 | 29 | 42 | 25 | 0.08 | -0.48 | 5.08 | -38.54 | 9.30 |
| Nguyen and Wilson (2009) | 21 | 3 | - | 18 | 17 | -1.10 | -1.83 | 2.08 | -8.22 | 0.12 |
| Olper and Raimondi (2008) | 11 | 5 | 1 | 6 | 5 | -0.43 | 0.68 | 5.97 | -8.66 | 12.00 |
| Otsuki et al. (2001a) | 25 | 23 | 3 | 2 | 1 | 0.88 | 1.25 | 1.38 | -0.91 | 5.20 |
| Otsuki et al. (2001b) | 2 | 2 | 2 | - | - | 0.74 | 0.74 | 0.44 | 0.43 | 1.05 |
| Péridy (2012) | 11 | - | - | 11 | 11 | -0.04 | -0.04 | 0.01 | -0.05 | -0.02 |
| Peterson et al. (2013) | 18 | 2 | - | 16 | 14 | -0.91 | -1.18 | 1.27 | -5.27 | 0.26 |
| Saitone (2012) | 40 | 24 | 1 | 16 | - | 0.06 | 0.05 | 0.43 | -1.18 | 1.77 |
| Scheepers et al. (2007) | 1 | 1 | 1 | - | - | 0.26 | 0.26 | - | 0.26 | 0.26 |
| Schlueter et al. (2009) | 40 | 23 | 11 | 17 | 8 | 0.02 | 0.00 | 0.36 | -0.87 | 0.88 |
| Schuster and Maertens (2013) | 68 | 25 | 3 | 43 | 33 | -0.08 | -0.08 | 0.16 | -0.64 | 0.27 |
| Shepherd and Wilson (2013) | 25 | 6 | 4 | 19 | 16 | -0.03 | -0.58 | 0.94 | -3.80 | 0.47 |
| Shepotylo (2016) | 72 | 35 | 9 | 37 | 8 | 0.00 | -0.04 | 0.23 | -0.82 | 0.71 |
| Sun et al. (2014) | 6 | 3 | - | 3 | - | -0.44 | -0.54 | 0.63 | -1.36 | 0.02 |
| Tran et al. (2014) | 52 | 50 | 43 | 2 | - | 0.18 | 0.99 | 1.21 | -0.01 | 3.54 |
| Vollrath et al. (2009) | 12 | 3 | - | 9 | 5 | -0.34 | -0.60 | 0.99 | -3.18 | 0.80 |

Table A.4. Descriptive statistics for papers included in the sample.

| References | Obs. | Positive obs. | Positive significant obs. | Negative obs. | Negative significant obs. | Me | µ | σ | Min | Max |
|---|---|---|---|---|---|---|---|---|---|---|
| Wei et al. (2012) | 20 | 16 | 9 | 4 | - | 0.04 | 0.05 | 0.09 | -0.09 | 0.21 |
| Wilson and Otsuki (2003) | 3 | 3 | 2 | - | - | 0.34 | 0.52 | 0.54 | 0.09 | 1.12 |
| Wilson and Otsuki (2004) | 3 | 3 | 3 | - | - | 1.45 | 1.42 | 0.07 | 1.34 | 1.48 |
| Wilson et al. (2003) | 2 | 2 | 2 | - | - | 0.59 | 0.59 | 0.01 | 0.58 | 0.59 |
| Winchester et al. (2012) | 28 | 14 | - | 14 | 7 | -0.08 | -1.14 | 3.22 | -9.87 | 2.82 |
| Xiong and Beghin (2011) | 24 | 11 | 3 | 13 | 5 | -0.01 | 0.34 | 0.98 | -0.72 | 3.00 |
| Xiong and Beghin (2014) | 6 | 3 | 3 | 3 | 3 | 0.23 | 0.24 | 0.61 | -0.42 | 0.93 |
| Yue and Beghin (2009) | 1 | 1 | 1 | - | - | 0.99 | 0.99 | - | 0.99 | 0.99 |